\documentclass[12pt]{article}
\usepackage{epsf}
\usepackage{amssymb}
\usepackage{latexsym}

\usepackage{graphicx,color}
\usepackage{wrapfig}
\usepackage{latexsym}
\usepackage{graphics}
\usepackage{amsmath} 

\usepackage{color}
\usepackage{delarray,color}
\usepackage{fancybox}  
\usepackage{tabularx}

\catcode`\@=11
\@addtoreset{equation}{section}
\def\theequation{\thesection.\arabic{equation}}
\catcode`@=12
\relax

\begin{document}
\baselineskip 0.7cm

\begin{titlepage}

\setcounter{page}{0}

\renewcommand{\thefootnote}{\fnsymbol{footnote}}
\vspace{-1cm}
\begin{flushright}
ITFA-2008-24
\end{flushright}

\vskip 1.35cm

\begin{center}
{\Large \bf 
The Volume Conjecture and Topological Strings
}

\vskip 1cm 

{\normalsize
Robbert Dijkgraaf$^{1}$\footnote{r.h.dijkgraaf@uva.nl}
and Hiroyuki Fuji$^{2}$\footnote{fuji@th.phys.nagoya-u.ac.jp}
}

\vskip 0.3cm

{ \it
$^1$ Institute for Theoretical Physics $\&$ Korteweg-de Vries Institute
 for Mathematics\\
University of Amsterdam, Valckenierstraat 65, 1018 XE Amsterdam, 
The Netherlands 
\\[2mm]
$^2$ Department of Physics, Nagoya University, Nagoya 464-8602, Japan
}

\end{center}

\vspace{12mm}

\centerline{{\bf Abstract}} In this paper, we discuss a relation
between Jones-Witten theory of knot invariants and 
topological open string theory on the basis of the volume conjecture. 
We find a similar Hamiltonian structure for both theories, 
and interpret the AJ conjecture as the ${\cal D}$-module structure 
for a D-brane partition function.
In order to verify our claim, we compute the free energy for 
the annulus contributions in the topological string using the 
Chern-Simons matrix model, and find that it coincides with 
the Reidemeister torsion in the case of
the figure-eight knot complement and the SnapPea census manifold
$m009$.

\end{titlepage}
\newpage

\section{Introduction}
In recent years, some remarkable progress in knot theory has been reported.
One of the most fascinating developments is the {\it volume conjecture} 
proposed by Kashaev \cite{Kashaev}.
In \cite{Murakami^2}, it is shown that 
the R-matrix for the
Kashaev's invariant and the colored Jones polynomial \cite{AD} 
are equivalent.
Thus, the volume conjecture can be expressed simply as
\begin{eqnarray}
\frac{1}{\pi}\lim_{n\to \infty}\frac{\log |J_n(K;q=e^{2\pi i/n})|}{n}
={\rm Vol}({\bf S}^3\backslash K),
\end{eqnarray} 
where $J_n(K;q)$ is an $n$-colored Jones polynomial for a hyperbolic 
knot $K$. 
This conjecture has been verified for various knots by analyzing the asymptotic 
behavior of the colored Jones polynomial \cite{KT}$\mathchar`-$\cite{Roland3} 
(For a comprehensive review, see \cite{HMurakami2}).
The volume conjecture has also been
generalized, and the complexified version is proposed in \cite{MMOTY}.
At first glance, the Jones polynomial does not seem to be related 
to the volume of the hyperbolic three-manifold. 
However, physically the claim is quite nautral 
from the point of view of the $SL(2;\mathbb{C})$ Chern-Simons gauge theory \cite{Witten1},
and the first-order formulation of the
three-dimensional gravity with a negative cosmological constant 
\cite{Witten2}.

Hyperbolic volumes are also interesting from the arithmetic 
point of view. 
In Boyd's work \cite{Boyd,BR1,BDR}, the relation between the hyperbolic 
volume and the logarithmic Mahler measure for the A-polynomials 
is studied analytically and numerically. Using the simplicial 
decomposition of the knot complement, we can express the hyperbolic
volume as the sum of Roger's dilogarithm functions related to 
the volumes of each ideal tetrahedra. 
In contrast, for some
hyperbolic knots, the logarithmic Mahler measures for the A-polynomials are
exactly evaluated and are given by a special value of an L-functions.
These two expressions of the hyperbolic volume give rise to nontrivial
identities, which are essentially captured by Bloch-Beilinson's
conjecture. 

A one-parameter extension of 
the volume conjecture is proposed in 
\cite{Gukov,Murakami-Yokota,HMurakami}. 
This version of the conjecture is called the
{\it generalized volume conjecture.}
Physically, the generalized volume conjecture implies the double scaling
limit for the level $k$ of the $SU(2)$ Chern-Simons gauge theory 
and the dimension $n$ of the representation for the Wilson loop along the knot. 
In this limit, the volume is 
generalized to the Neumann-Zagier's potential function 
\cite{NZ}, which describes 
the deformation of the complete structure of the knot complement.
In this paper we discuss some correspondences between the 
Jones-Witten theory and topological open string theory, which can be
regarded as a geometric engineering of the $SU(2)$ 
Chern-Simons gauge theory.
We propose a correspondence between the colored Jones
polynomial and the partition function for the topological open string
on the basis of the generalized volume conjecture.

In the topological B-model, the free energy for the disk contributions 
\cite{OV} is given by an Abel-Jacobi map 
on a holomorphic curve inside a non-compact Calabi-Yau threefold 
\cite{AV,AKV}. 
In contrast, the Neumann-Zagier's function is also given 
by an analytic continuation of the Abel-Jacobi map on 
the character variety.
Futhermore, a conjecture on
the difference equation for the colored Jones polynomial is
proposed \cite{Garou1,Garou2,Garou-Le}.
This difference equation is quite analogous to the ${\cal D}$-module structure of 
the partition function of the topological open string.
This analogy is consistent with the volume conjecture; thus, we expect the 
following relation to hold under some appropriate analytic 
continuation.
\begin{eqnarray}
J_n(K;q)\simeq Z_{\rm open}(u;q),
\end{eqnarray}
where the topological open string is defined on  
the Calabi-Yau threefold $X^{\vee}$,
\begin{eqnarray}
X^{\vee}:=\left\{(x,y,z,w)\in \left(
\mathbb{C}^{*}\right)^2\times \mathbb{C}^{2}
\bigl|
zw={A}_K(x,y)
\right\}.
\end{eqnarray}
$A_K(x,y)$ is an A-polynomial \cite{CCGLS} for
the knot $K$.
In the subleading order of the WKB expansion of both sides this
correspondence implies that
the Reidemeister torsion on the knot
complement corresponds to  the annulus free energy in the topological
string. 
In this paper, we will check this conjecture for the figure-eight knot
complement and the SnapPea census manifold $m009$.

The organization of this paper is as follows:
In section 2, we review the volume conjecture and the AJ conjecture.
In section 3, we discuss the topological open string theory. 
In particular, we discuss the analogy in the computation of the free
energy for the disk contributions and the ${\cal D}$-module structure of 
the open string partition function.
In section 4, we compute the annulus free energy using the 
Chern-Simons matrix model and find that it equals the 
Reidemeister torsion.
In Appendix,  we summarize the derivation of the open string free
energies using the large $N$ analysis of the Chern-Simons 
matrix model.

\section{Review of volume conjecture and AJ conjecture}
\subsection{Hyperbolic three-manifold and volume}
Let $M$ be a hyperbolic three-manifold with a finite volume.
In general, such a three-manifold is constructed as the quotient of the
three dimensional hyperbolic space ${\bf H}^3$. 
The metric for ${\bf H}^3$ defined on the upper half space is given by
\begin{eqnarray}
ds^2=\frac{dx^2+dy^2+dz^2}{z^2}, \quad (x,y)\in \mathbb{R}^2, \;\;
z\in \mathbb{R}_+.
\label{metric}
\end{eqnarray}
The isometry group of ${\bf H}^3$ is 
$PSL(2;\mathbb{C})$. It acts as 
\begin{eqnarray}
&&q\to q^{\prime}=\frac{aq+b}{cq+d}, \quad 
\left(
\begin{array}{cc}
a & b \\
c & d
\end{array}
\right)\in PSL(2;\mathbb{C}),
\label{isom}
\\
&&q:=x+{\bf i}y+{\bf j}z.
\end{eqnarray}
The hyperbolic three-manifold $M$ is given by
\begin{eqnarray}
M={\bf H}^3/\Gamma,
\end{eqnarray}
where $\Gamma$ is a torsion-free and discrete subgroup of $PSL(2;\mathbb{C})$
with the action (\ref{isom}).

Topologically, the hyperbolic three-manifolds can also be constructed 
as a knot complement space \cite{Thurston}
\begin{eqnarray}
M={\bf S}^3\backslash N(K),
\end{eqnarray}
where $N(K)$ is the tubular neighbourhood of a knot $K$.
Hereafter, we shall denote the knot complement
as ${\bf S}^3\backslash K$.
\begin{figure}[h]
\hspace*{5cm}
\includegraphics[width=6cm,clip]{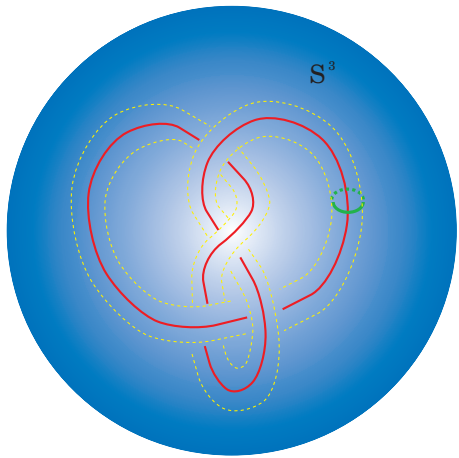}
\caption{Knot complement}
\end{figure}
A hyperbolic structure is not admitted for every knot complement
space. In fact, the complement of torus knots and satellite knots
does not admit a hyperbolic structure.
The knots that admit the hyperbolic structure for 
their ${\bf S}^3$ complement are called {\it hyperbolic knots}.
In the celebrated work of Thurston, it is shown that all knots can be 
classified as torus knots, satellite knots, and hyperbolic knots.
In this paper, we mainly consider the hyperbolic knot complements.

If a three-manifold admits a hyperbolic structure, 
the three-manifold can be decomposed simplicially into 
{\it ideal tetrahedra}. The vertices of an ideal tetrahedron are located 
at conformal infinity, and the edges are geodesics connecting each vertex 
with respect to the metric (\ref{metric}). There are six face angles 
along each edge, and the face angles on the opposite edges have 
the same value. Therefore, an ideal tetrahedron is specified by 
three face angles, $\alpha$, $\beta$, and $\gamma$, that satisfy 
the ideal triangle condition $\alpha+\beta+\gamma=\pi$.
\begin{figure}[h]
\hspace*{2cm}
\includegraphics[width=12cm,clip]{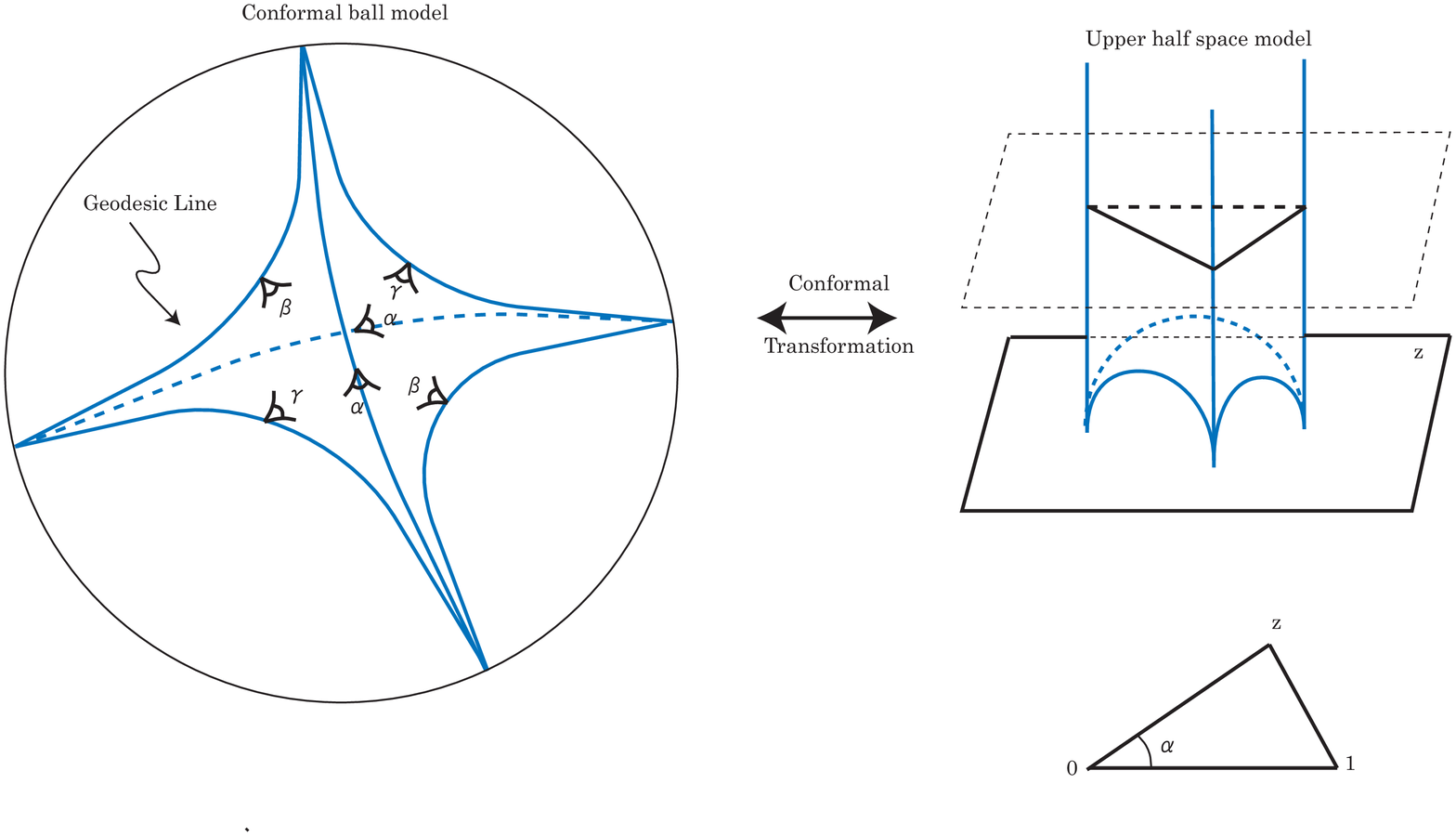}
\caption{Ideal tetrahedron}
\end{figure}

The volume of an ideal tetrahedron $T_{\alpha\beta\gamma}$ 
is computed directly by using the metric (\ref{metric}) \cite{Thurston}.
\begin{eqnarray}
&&{\rm Vol}(T_{\alpha\beta\gamma})
=\Lambda(\alpha)+\Lambda(\beta)+\Lambda(\gamma),
\label{ideal volume} \\
&&\Lambda(\theta):=-\int_0^{\theta}dt\;\log |{\sin 2t}|.
\end{eqnarray}
The function $\Lambda(\theta)$ is called the Lobachevsky function.
The volume formula can also be rewritten in terms of the Bloch-Wigner 
function $D(z)$
\begin{eqnarray}
&& {\rm Vol}(T_{\alpha\beta\gamma})=D(z),
\quad {\rm arg}(z_1)=\alpha,\;{\rm arg}(z_2)=\beta,\;
{\rm arg}(z_3)=\gamma,\\
&& D(z):={\rm Im}\; Li_2+{\rm arg}(1-z)\log |z|,
\quad 
z_1=z,\; z_2=1-\frac{1}{z},\;z_3=\frac{1}{1-z}.
\end{eqnarray}

The gluing condition along each edge imposes constraints
on the face angles of the ideal tetrahedra.
\begin{eqnarray}
\prod_{\rm edge}z_i=1. 
\label{edge}
\end{eqnarray}
Furthermore, we should also consider the gluing condition for the
boundary of the knot complement.
In order to realize the torus as a boundary, the face angles must satisfy
the completeness conditions along the meridian and the longitude.
\begin{eqnarray}
\prod_{\rm meridian}z_i=1,\quad 
\prod_{\rm longitude}z_i=1.
\label{complete1}
\end{eqnarray} 
Mostow's rigidity theorem \cite{Mostow}
implies that these conditions are solved 
uniquely with ${\rm Im}\; z_i>0$ for all $i$. 
By summing all volumes for the ideal tetrahedra,
we are able to determine
the hyperbolic volume uniquely for each hyperbolic
three-manifold. 
\begin{figure}[h]
\hspace*{1cm}
\includegraphics[width=15cm,clip]{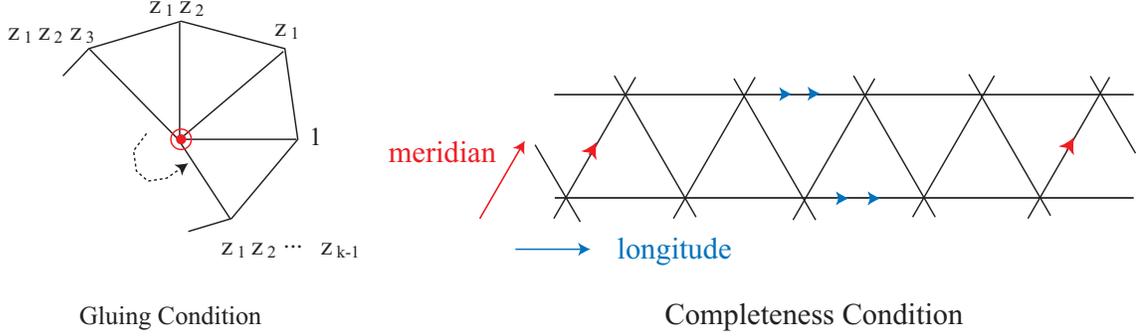}
\caption{Gluing condition along each edge and completeness condition}
\end{figure}

\subsection{A-polynomial and logarithmic Mahler measure}
The holonomy representations along the meridian $\mu$ and the 
longitude $\nu$ of the boundary torus are given in general by
\begin{eqnarray}
&& \rho(\mu)=\left(
\begin{array}{cc}
m^{1/2} & * \\
0 & m^{-1/2}
\end{array}
\right), \quad 
\rho(\nu)=\left(
\begin{array}{cc}
\ell & * \\
0 & \ell^{-1}
\end{array}
\right), \quad
 (m,\ell)\in \mathbb{C}^{*}\times \mathbb{C}^{*}.
\label{holonomy}
\end{eqnarray}
Since the holonomy obeys the relation of the knot group 
$\pi_1(K):=\pi_1({\bf S}^3\backslash K)$,
the eigenvalues $m$ and $\ell$ must be a solution of  
an algebraic equation \cite{CCGLS}
\begin{eqnarray}
A_{K}(m,\ell)=0.
\label{A-polynomial}
\end{eqnarray}
The polynomial $A_{K}(m,\ell)$ is called the {\it A-polynomial}, and the
character variety $X_K$ is defined by
\begin{eqnarray}
X_K:=\{(m,\ell)\in \mathbb{C}^{*}\times \mathbb{C}^{*}|
A_{K}(m,\ell)=0
\}.
\end{eqnarray}

There are many remarkable properties of A-polynomials.
Three of these properties are listed as follows:
\begin{enumerate}
\item Reciprocal: $A_K(m,\ell)=\pm A_K(1/m,1/\ell)$.
\item Tempered: All edge polynomials of $A_K(m,\ell)$ are cyclotomic.
\item $SL(2;\mathbb{Z})$ transformation: 
If one changes the homotopy basis $\mu$, $\nu$ 
of the boundary torus  
\begin{eqnarray}
\left(
\begin{array}{c}
\nu \\
\mu
\end{array}
\right)
\to
\left(
\begin{array}{cc}
a & b \\
c & d
\end{array}
\right)
\left(
\begin{array}{c}
\nu\\
\mu
\end{array}
\right),
\quad 
\left(
\begin{array}{cc}
a & b \\
c & d
\end{array}
\right)\in 
SL(2;\mathbb{Z}),
\end{eqnarray}
the A-polynomial transforms as
\begin{eqnarray}
A_K(m,\ell)\to A_K(m^a\ell^{-c},m^{-b}\ell^d),
\end{eqnarray}
up to the factor $\pm m^e\ell^f$ ($e,f\in \mathbb{Z}$).

\end{enumerate}

In \cite{Boyd,BR1,BDR}, 
the relation between the volume of the 
hyperbolic knot complement and 
the logarithmic Mahler measure of the A-polynomial 
is observed numerically.
From the analytical and the numerical studies, a simple relation is
found for some hyperbolic knots,
\begin{eqnarray}
{\rm Vol}({\bf S}^3\backslash K)=\pi m(A_K),
\end{eqnarray}
where $m(P)$ is a {\it logarithmic Mahler measure} for a two-parameter polynomial
$P(x,y)$.
\begin{eqnarray}
m(P)&:=&\int_{|x|=1}\frac{dx}{2\pi i}\int_{|y|=1}\frac{dx}{2\pi i}
\log|P(x,y)|
\nonumber \\
&=&\sum_{i=1}^{n}\int_{|x|=1}\frac{dx}{2\pi i}\log^+ |y_i(x)|
\end{eqnarray}
In the second line, we used Jensen's formula \cite{Ahlfols}. 
The function $\log^+ |z|$ is $\log^+ z=\log z$ for $|z|\ge 1$
and vanishes otherwise. $y_i(x)$ are the solutions of $P(x,y)=0$.

\subsection{Generalized volume conjecture}
A small deformation of the completeness condition 
is studied in the work of Neumann and Zagier \cite{NZ}.
They considered the deformation of the completeness condition to be
\begin{eqnarray}
\prod_{\rm meridian}z_i=m, \quad \prod_{\rm longitude}z_i=\ell^2.
\label{deformation}
\end{eqnarray}
The deformation parameters are 
identified with the eigenvalues of the holomony matrices $\rho(\mu)$ 
and $\rho(\nu)$. The holonomy representation becomes reducible if 
$m,\ell=1$. 
Therefore, the complete point should correspond to $(m,\ell)=(1,-1)$.
For later convenience, we introduce the parameters $(u,v)$\footnote{
Our definition of $(u,v)$ is related to $(\mathfrak{u},\mathfrak{v})$ in \cite{CCGLS,GM} 
and $(\tilde{u},\tilde{v})$ in \cite{Gukov} as follows:
\begin{eqnarray}
u=\mathfrak{u}=2\tilde{v}, \quad 2v=\mathfrak{v}=-2\tilde{u}.
\end{eqnarray}
}
\begin{eqnarray}
u:=\log(m),\quad v:=\log(\ell)+\pi i.
\end{eqnarray}

Let $M_{u}$ be the incomplete manifold with the boundary holomomy
representations $\rho(\mu)$ and $\rho(\nu)$.
The incomplete manifold can be completed via $(p,q)$-Dehn surgery, where 
$(p,q)$ are coprime integers and satisfy $pu+2qv=2\pi i$.
The volume and Chern-Simons functions \cite{CS} for $M_{u}$ are
given in \cite{NZ,Yoshida,KK1,KK2} as
\begin{eqnarray}
&& {\rm Vol}(M_{u}):={\rm Vol}(M)+{\rm Im}\;f(u)
+\frac{1}{2}{\rm Im}(u\bar{v}(u)),\\
&& {\rm CS}(M_{u}):={\rm CS}(M)-{\rm Re}\;f(u)
+\frac{\pi}{2} {\rm Im}(ru+2sv), \quad {\rm mod}\; \pi^2,
\end{eqnarray}
where the coprime integers $(r,s)$ satisfy $ps-qr=1$.
The potential function $f(u)$ is determined by Schl\"afli's formula
for the variation of the face angles in the ideal tetrahedra 
\cite{Hodgson,Dunfield}. 
\begin{eqnarray}
\frac{df(u(t))}{dt}=-\left[
{\rm Re}\bigl(u(t)\bigr)\frac{d{\rm Im}\bigl(v(t)\bigr)}{dt}
-{\rm Re}\bigl(v(t)\bigr)\frac{d{\rm Im}\bigl(u(t)\bigr)}{dt}
\right],
\end{eqnarray}
where $(u(t),v(t))$ satisfies $A_K(e^{u(t)},e^{v(t)})=0$.
This integration path ${\cal L}_u$, which connects the complete point
$(0,\pi i)$ and the deviation point $(u,v+\pi i)$,
is the Lagrangian submanifold in the character variety $X_K$.

On the basis of this deformation of the hyperbolic structure, 
a one-parameter extension of the volume conjecture has been proposed 
\cite{Gukov,Murakami-Yokota,HMurakami}.
This extended conjecture is called the {\it generalized volume conjecture}.
The claim of the generalized volume conjecture is summarized 
as follows:
\begin{eqnarray}
&& 2\pi \lim_{n,k\to \infty}\frac{J_n(K;e^{2\pi i/k})}{k}
=H(u), \quad u:=2\pi i\left(\frac{n}{k}-1\right).
\end{eqnarray}
The function $H(u)$ is called the {\it Neumann-Zagier's potential function}
\cite{Murakami-Yokota}, and it is related with $f(u)$ as follows:
\begin{eqnarray}
H(u)-H(0)
=-f(u)
+\pi i u+\frac{1}{2}uv.
\end{eqnarray}
The essential part of the function $H(u)$
is given by the integrations
\begin{eqnarray}
&&H(u)\sim {\rm Vol}(m,\ell)+2\pi^2i{\rm CS}(m,\ell).
\\
&& {\rm Vol}(m,\ell):= {\rm Vol}({\bf S}^3\backslash K)
+\int_{{\cal L}_u}[-\log|\ell|d({\rm arg}\;m)
+\log|m|d({\rm arg}\;\ell)],
\label{volume integral}\\
&& {\rm CS}(m,\ell):=\frac{1}{2\pi^2}
{\rm CS}({\bf S}^3\backslash K)
-\frac{1}{2\pi^2}\int_{{\cal L}_u}[\log|m|d\log|\ell|
+({\rm arg}\;m) d({\rm arg}\;\ell)].
\end{eqnarray}
The parameters 
$m=e^{u}$ and $\ell=-e^{v}$ satisfy $A_{K}(m,\ell)=0$.
In order to match the above result with the result of the Dehn filling, we need to add the 
linear linear terms of $u$ and $v$.\footnote{This modification becomes
relevant for the case  ${\rm Re}\;u\ne 0$.}
From the Schl\"afli formula, $H(u)$ satisfies
\begin{eqnarray}
v(u)+\pi i=\frac{dH(u)}{du}.
\label{canonical relation}
\end{eqnarray}
On the basis of the semi-classical analysis of the Chern-Simons gauge theory,
this relation can be naturally interpreted in terms of the 
canonical coordinate $\log m$, momentum $\log \ell$, 
and Hamiltonian $A_K$. Such physical aspects will be discussed in
section 2.4.

Since the volume is given by the logarithmic Mahler measure 
for some class of knots,
the volume term is also given by the formula (\ref{volume integral})
with the integration path on $|m|=1$ with $|\ell|\ge 1$.
By utilizing the reciprocal and tempered properties of A-polynomial,
we can also include the volume term in the volume function by
extending the integration path ${\cal L}_u$ appropriately. 

In \cite{Gukov,GM}, the volume conjecture is further extended to 
the subleading order: 
\begin{eqnarray}
&&J_n(K,e^{\frac{2\pi i}{k}})\underset{\tiny k,l\to \infty}{\sim}
\frac{k}{2\pi}H(u)+\frac{1}{2}\delta_K(u)\log k
+\frac{1}{2}\log\left(
\frac{T_K(u)}{2\pi^2}
\right)
+\sum_{n=1}^{\infty}\left(\frac{2\pi}{k}\right)^nS_{n+1}(u_0),
\nonumber \\
&&\label{subleading}
\end{eqnarray}
where $\delta_K(u)$ is a number that is determined by the topology of
the knot complement and the representation $\rho$. 
The function $T_K(u)$ is the {\it Reidemeister torsion}
$T({\bf S}^3\backslash K,\rho)$ of the knot complement
twisted by the flat connection corresponding to the representation 
$\rho$.  

\noindent\underline{\it Definition of Reidemeister torsion:}

In \cite{Milnor,Turaev}, 
the Reidemeister torsion for the chain complex
\begin{eqnarray}
C_*:0\to C_n\stackrel{\partial_n}{\to}C_{n-1}\stackrel{\partial_{n-1}}{\to}\cdots\stackrel{\partial_1}{\to}C_0\to 0
\end{eqnarray}
is defined. 
The image $B_i:={\rm im}\{d_{i+1}:C_{i+1}\to C_{i}\}$,  
the kernel $Z_i:={\rm ker}\{d_{i}:C_{i}\to C_{i-1}\}$, 
and the homology group $H_i:=Z_i/B_i$
are defined for each $i$.
There exist the exact sequences
\begin{eqnarray}
 0\to Z_i\to C_i\to B_{i-1}\to 0
\end{eqnarray}
and
\begin{eqnarray}
 0\to B_i\to Z_i\to H_i\to 0.
\end{eqnarray}
In particular, the maps $s_i:B_i\to C_{i+1}$ and 
$\bar{s}_i:H_i\to Z_i$ play important roles in the costruction of the basis.

Let ${\bf c}^i$, ${\bf b}^i$, and ${\bf h}^i$ be 
the reference bases of $C_i$, $B_i$, and $H_i$, respectively.
Using the bases ${\bf b}^i$ and ${\bf h}^i$, we find that the complete basis of 
$C_i$ is spanned by ${\bf b}^i\sqcup s_{i-1}({\bf b}^{i-1})\sqcup
\bar{s}_{i}({\bf h}^{i})$. 
Then, the Reidemeister torsion of the chain complex $C_*$ is 
the alternating product
\begin{eqnarray}
{\rm tor}(C_*,{\bf c}^*,{\bf h}^{*})=\prod_{i=1}^{n}[{\bf b}^i\sqcup s_{i-1}({\bf b}^{i-1})\sqcup
\bar{s}_{i}({\bf h}^{i})/{\bf c}^i]^{(-1)^{i+1}}.
\end{eqnarray}
In the above expression, we denote $[{\bf f}/{\bf g}]$ for the ordered basis 
${\bf f}:=\{f_1,\cdots ,f_n\}$ and ${\bf g}:=\{g_1,\cdots ,g_n\}$ with 
$f_i=\sum_j p_{ij} g_j$ as $[{\bf f}/{\bf g}]:=\det (p_{ij})$.  

The Reidemeister torsion of three-manifold $M$ is given by the finite
CW-complex $K$ for $M$. Let $X$ and $\tilde{X}$ be a finite cell 
complex and its universal covering. The fundamental group $\pi_1(X)$ 
acts on $\tilde{X}$ as the deck transformation.
The chain complex $C_*(\tilde{X};\mathbb{Z})$ has the left
$\mathbb{Z}[\pi_1(X)]$-module structure.  
Using a $SL(2;\mathbb{C})$ representation $\rho$,
we can express the $sl(2;\mathbb{C})_{\rho}$-twisted chain complex 
for the CW-complex $K$ with the underlying topological space 
$|K|=X$ as
\begin{eqnarray}
C_*(K;Ad_{\rho})
=sl(2;\mathbb{C})\otimes_{\mathbb{Z}[\pi_1(X)]}C_*(\tilde{K};\mathbb{Z}),
\end{eqnarray}
where the action of $Ad\rho$ is $Ad\rho(\gamma,g):=Ad_{\rho(\gamma)}g\in
sl(2:\mathbb{C})$ for $(\gamma,g)\in(\pi_1(X),sl(2;\mathbb{C}))$.

For this twisted chain complex, the Reidemeister torsion 
${\rm tor}(K;Ad\rho,h^i)$
is defined naturally.
\begin{eqnarray}
{\rm tor}(K;Ad\rho,\{h^i\})
:={\rm tor}(C_i(K;Ad\rho),\{{\cal A}\otimes c^i\},\{h^i\}).
\end{eqnarray}
In order to extend the basis for the twisted chain complex, we introduced 
${\cal A}$ as a $\mathbb{C}$-basis of $sl(2;\mathbb{C})$
and $h^i$ as a $\mathbb{C}$-basis of $H_i(K;Ad\rho)$.

For the three-dimensional manifold $M$, the Reidemeister
torsion is computed as \cite{Porti,Dubois}
\begin{eqnarray}
T(M,\rho):=\exp\left(
-\frac{1}{2}\sum_{n=0}^{3}n(-1)^n\log \det {}^{\prime}\Delta_n^{E_\rho}
\right),
\end{eqnarray}
where $\Delta_n^{E_\rho}$ is a Laplacian on $n$-forms with the
coefficients in the flat bundle $E_{\rho}$. $T(M,\rho)$ is a torsion 
${\rm tor}(M;Ad\rho,\{h^1,h^2\})$ with a sign refinement.
In particular, for a knot complement, we denote the Reidemeister torsion 
$T_K(u):=T({\bf S}^3\backslash K,\rho(\mu))$.

\subsection{Physical derivation of volume conjecture}
In \cite{Gukov,Dinesh}, the generalized volume conjecture is 
derived physically.
In terms of the
$SU(2)$ Chern-Simons gauge theory, 
the Jones polynomial $J_n(K;q)$ is given by 
a vacuum expectation value of the Wilson loop operator 
$W_R(K;q)$ along the knot $K$ on ${\bf S}^3$ \cite{Witten1}.
\begin{eqnarray}
W_R(K;q):={\rm Tr}_{R}{\rm P}\exp\left(i
\oint_K A
\right),
\end{eqnarray}
where the representation of $SU(2)$ is chosen as $R=(n)$.

According to the geometric quantization scheme \cite{Murayama},
the holonomy around the Wilson loop for the $SU(2)$ Chern-Simons gauge theory, 
is found. 
In the following,
we briefly review the derivation in \cite{Murayama}
and find the correct normalization of $(u,v)$ in Chern-Simons gauge theory.

In order to perform the Hamiltonian quantization 
we choose the time direction to be along the knot.
The expectation value of the Wilson loop is given in terms of the path
integral
\begin{eqnarray}
\langle
W_R(K)
\rangle
=\int {\cal D}A\;{\rm Tr}_R \;{\rm T}\exp 
\Bigl[i\int_K dt\;A_0(x)\Bigr]{\rm e}^{iS_{\rm CS}[A]},
\end{eqnarray}
where T denotes the time ordering.
The quantization of the gauge fields with the Wilson loop operator is not
straightforward, since $W_R(K)$ cannot be exponentiated because of
the trace with respect to the representation $R$.

In order to overcome this point, we take a trace after the quantization 
by making use of the Borel-Weil theory \cite{Woodhouse}.
This theory implies that the representation space of the
group $G$ is isomorphic to the space of the holomorphic sections on
$G/T$, where $T$ is the maximal torus of $G$. 
In the case of $G/T=SU(2)/U(1)\simeq {\bf P}^1$,
the holomorphic section is given by a complex scalar field $\zeta(t)$.
It is specified by the 
transition function between ${\bf P}^1\backslash \{0\}$
and ${\bf P}^1\backslash \{\infty\}$.
If the transition function is chosen to be $\zeta^{-2j}$,
the space of the holomorphic section is spanned by $\zeta^n$
($n=0,\cdots, 2j$) and its dimension is $2j+1$.

In the geometric quantization, the $SU(2)$ generators 
acting on this space of the holomorphic section are represented by the 
symplectic form $\omega\in H^2({\bf P}^1)$.
For the spin $j$ representation, the symplectic form is
\begin{eqnarray}
\omega=\frac{1}{i}\frac{2jd\bar{\zeta}\wedge d\zeta}{(1+\zeta\bar{\zeta})^2}.
\label{symp BW}
\end{eqnarray}
From this symplectic form the canonical transformations of the 
Killing vectors for $SU(2)/U(1)$ are
\begin{eqnarray}
&& J^+[\zeta,\bar{\zeta}]=j\frac{2\bar{\zeta}}{1+z\bar{\zeta}}, 
\quad J^{-}[\zeta,\bar{\zeta}]=j\frac{2\zeta}{1+\zeta\bar{\zeta}},
\quad J^3[\zeta,\bar{\zeta}]=j\frac{1-\zeta\bar{\zeta}}{1+\zeta\bar{\zeta}}.
\end{eqnarray}
By using these representations, we can rewrite
the expectation value of the Wilson loop operator as 
\cite{Murayama}
\begin{eqnarray}
\int {\cal D}A\;{\rm Tr}_R \;{\rm T}\exp 
\Bigl[i\int_K dt\;A_0(x)\Bigr]{\rm e}^{iS_{\rm CS}[A]}
=\int {\cal D}A{\cal D}{\zeta}{\cal D}\bar{\zeta}e^{iS_{rm CS}[A]+iS_{G/T}[\zeta]}
e^{S_{\rm WL}[\zeta,A_0]},
\end{eqnarray}
where $S_{\rm G/T}[\zeta]$ and $S_{\rm WL}[\zeta,A_0]$ are
\begin{eqnarray}
S_{\rm G/T}[\zeta]:=\frac{1}{i}\int dt\;\frac{2j\bar{\zeta}\dot{\zeta}}{1+\zeta\bar{\zeta}},\quad
S_{\rm WL}[\zeta,A_0]:=\int dt \; {\rm Tr}A_0(t)\sum_{i=1}^3 J^i[\zeta,\bar{\zeta}] \sigma_i.
\end{eqnarray}
Thus, we find the Gauss' law constraint 
\begin{eqnarray}
\frac{k}{4\pi}F_{12}=\delta^2(z-w)
\frac{j}{1+\zeta\bar{\zeta}}\left(
\begin{array}{cc}
1-\zeta\bar{\zeta} & 2\bar{\zeta} \\
2\zeta & \zeta\bar{\zeta}-1
\end{array}
\right).
\end{eqnarray}
In the holomorphic gauge $A_{\bar{z}}=0$, the gauge field $A_{z}$
is solved explicitly.
\begin{eqnarray}
A_z=\frac{1}{k}\frac{1}{z-w}
\frac{j}{1+\zeta\bar{\zeta}}\left(
\begin{array}{cc}
1-\zeta\bar{\zeta} & 2\bar{\zeta} \\
2\zeta & \zeta\bar{\zeta}-1
\end{array}
\right).
\end{eqnarray}
From this solution, we can compute the holonomy around the
Wilson loop to find
\begin{eqnarray}
{\rm P}\exp \Bigl[
i\oint_w A_z dz
\Bigr]=\cos\frac{2\pi j}{k}\mathbb{I}_2+i\sin\frac{2\pi j}{k}
\frac{1}{1+\zeta\bar{\zeta}}\left(
\begin{array}{cc}
1-\zeta\bar{\zeta} & 2\bar{\zeta} \\
2\zeta & \zeta\bar{\zeta}-1
\end{array}
\right).
\end{eqnarray}
This holonomy matrix is diagonalized as
\begin{eqnarray}
U^{\dagger}{\rm P}\exp \Bigl[
i\oint_w A_z dz
\Bigr]U
=\left(
\begin{array}{cc}
e^{2\pi ij/k} & 0 \\
0 & e^{-2\pi ij/k}
\end{array}
\right).
\label{Wilson hol}
\end{eqnarray}
By comparing this solution with the holonomy of the knot complement
(\ref{holonomy}), we can read off the parameter $u=\log m$ as
\begin{eqnarray}
u\equiv \frac{4\pi ij}{k}= \frac{2\pi in}{k}-\frac{2\pi i}{k},\quad
{\rm mod}\;2\pi i,
\end{eqnarray}
where $n:=2j+1$ denotes the dimension of the spin $j$ representation.
In the large $k$ limit the second term can be neglected.

In the basis of the axioms of the topological field theory \cite{Atiyah},
the expectation value of the Wilson loop operator can be decomposed into 
the knot complement and the solid torus with a Wilson loop
\begin{eqnarray}
\langle W_R(K)\rangle=
\int {\cal D}{\cal A}\; Z({\bf S}^3\backslash K)[{\cal A}]\cdot
\langle
 W_R(U)\rangle_{{\bf S}^1\times {\bf D}^2}[{\cal A}],
\end{eqnarray}
where ${\cal A}$ is the gauge field on the torus $\partial ({\bf S}^3\backslash
K)=\partial ({\bf S}^1\times {\bf D}^2)$.
Since the expectation value of the Wilson loop operator inside the solid
torus is known to give a delta function \cite{Dinesh,Moore-Seiberg},
this integration can be performed directly  
and we find that
\begin{eqnarray}
\langle W_R(K)\rangle=
\int {\cal D}\tilde{u}\; Z({\bf S}^3\backslash K;\rho(\mu)_{\tilde{u}})\cdot
\delta\left(\tilde{u}-\frac{4\pi ij}{k}\right)
=Z({\bf S}^3\backslash K;\rho(\mu)).
\end{eqnarray}
Thus, we find that the expectation value of the Wilson loop operator is 
equivalent to the partition function of the knot complement with 
a specific holonomy defined by the representation and the coupling
constant of the Chern-Simons gauge theory.

For the complexified the coupling constant $k$ of $SU(2)$ Chern-Simons
gauge theory, we obtain  the $SL(2;\mathbb{C})$ Chern-Simons gauge theory 
\begin{eqnarray}
I=\frac{t}{8\pi}\int_M{\rm Tr}\left(
A\wedge dA+\frac{2}{3}A\wedge A\wedge A
\right)
+\frac{\bar{t}}{8\pi}\int_M{\rm Tr}\left(
\bar{A}\wedge d\bar{A}+\frac{2}{3}\bar{A}\wedge \bar{A}\wedge \bar{A}
\right),
\end{eqnarray}
where $t:=k+i\sigma$ is a complexified coupling and $A$ and $\bar{A}$ are 
the $SL(2;\mathbb{C})$ gauge fields.
In \cite{Witten2}, it is found that the $SL(2;\mathbb{C})$ Chern-Simons
gauge theory is equivalent to the first-order formulation of the
three-dimensional gravity with a negative
cosmological constant, if one identifies the dreibein $e$ and spin
connection $w$ as
\begin{eqnarray}
A=w+ie, \quad \bar{A}=w-ie.
\end{eqnarray}
By expanding the action of the $SL(2;\mathbb{C})$ Chern-Simons gauge theory,
we obtain the first ordered form of the action for the
gravity with a topological term 
\begin{eqnarray}
I[e,w]&=&
-\frac{\sigma}{2\pi}\int_M {\rm Tr}\left(
w\wedge de+w\wedge w\wedge e-\frac{1}{3}e\wedge e\wedge e
\right)
\nonumber \\
&&
+\frac{k}{4\pi} \int_M
{\rm Tr}\left(w\wedge dw-e\wedge de
+\frac{2}{3}w\wedge w\wedge w-2w\wedge e\wedge e
\right).
\end{eqnarray}
The equation of motion gives rise to the Einstein equation
 $R_{ij}=-2g_{ij}$ which shows that the on-shell geometry is 
a hyperbolic three-manifold.

The partition function for three-dimensional gravity
on $M$ factorizes
holomorphically.\footnote{
In \cite{WittenECFT}, a novel proposal for three dimensional gravity is
conjectured. Here, we only consider the semi-classical aspects \cite{Maloney}: 
therefore our discussion will not contradict to the extremal CFT proposal.
}
\begin{eqnarray}
Z_{\rm grav}(M;q)=|Z_{SU(2)\;{\rm CS}}(M;q)|^2.
\end{eqnarray}
The partition function $Z_{\rm grav}(M_u;q)$
is expanded semi-classically.
\begin{eqnarray}
Z_{\rm grav}(M_u;q)=\exp\left[\frac{i\sigma}{2\pi}{\rm Vol}(M_{u})+\pi ik{\rm
 CS}(M_{u})+\cdots\right].
\label{gravity partition}
\end{eqnarray}

We analyze 
the leading terms in the gravity partition function
in terms of the WKB quantization of the
$SL(2;\mathbb{C})$ Chern-Simons gauge theory. 
The classical moduli space is the space of the flat $SL(2;\mathbb{C})$ gauge
connection $F=\bar{F}=0$ on $M$ modulo gauge equivalence.
As is well known, the flat connection is determined by the holonomy
representation
\begin{eqnarray}
X_M={\rm Hom}_{\mathbb{C}}(\pi_1(M);SL(2;\mathbb{C}))/SL(2;\mathbb{C}).
\end{eqnarray}
Thurston showed that  
${\rm Hom}_{\mathbb{C}}(\pi_1(M);SL(2;\mathbb{C}))$ is four-dimensional, 
if $M$ is a hyperbolic three-manifold.
In the case of the knot complement $M={\bf S}^3\backslash K$,
the moduli space $X_M$ of the flat connection coincides with 
the definition of
the character variety $X_K$ \cite{CCGLS}.

The quantization of $(u,v)$ on $X_K$ is found as follows 
\cite{NRZ,Gukov}:
In the temporal gauge $A_0=0$, the Poisson brackets for the gauge fields
$A$ and $\bar{A}$ are given by 
\begin{eqnarray}
&&\{A_i^a(x),A_j^b(y)\}_{\rm PB}
=\frac{4\pi}{t}\delta^{ab}\epsilon_{ij}\delta^2(x-y),
\quad 
\{\bar{A}_i^a(x),\bar{A}_j^b(y)\}_{\rm PB}
=\frac{4\pi}{\bar{t}}\delta^{ab}\epsilon_{ij}\delta^2(x-y),
\nonumber 
\\
&& \{\bar{A}_i^a(x),A_j^b(y)\}_{\rm PB}=0.
\label{PB gauge}
\end{eqnarray}
The coordinates $(m,\ell)$ are eigenvalues of the holomomy 
along the boundary torus (\ref{holonomy}).
Since the meridian and longitude cycles intersect at one point on the
torus, the Poisson bracket relation of $u:=\log m$ and $v:=\log\ell$ 
yields\footnote{
Here, we  use the normalization of the generators 
${\rm Tr}T^aT^b=\delta^{ab}$. Then, the factor $\delta^{ab}$ in (\ref{PB gauge}) 
gives rise to a factor $T^3T^3=\frac{1}{2}\mathbb{I}$. 
Because of this factor $1/2$, the Poisson bracket relation (\ref{PB uv}) for $u$ 
is obtained.
}
\begin{eqnarray}
\{u,v\}_{\rm PB}=\frac{4\pi}{t}, \quad 
\{\bar{u},\bar{v}\}_{\rm PB}=\frac{4\pi}{\bar{t}}.
\label{PB uv}
\end{eqnarray}
In particular, for the case $k=-i\sigma$, the chiral part of the Poisson
bracket gives
\begin{eqnarray}
\{u,v\}_{\rm PB}=\frac{2\pi}{k}.
\label{commutation}
\end{eqnarray}
From the above relation, we can read off the symplectic form. 
Utilizing the geometric quantization scheme \cite{Woodhouse},
we can express the semi-classical value of the action of $SL(2:\mathbb{C})$ 
Chern-Simons gauge theory as the phase function $S$
\begin{eqnarray}
&& S=\int_{{\cal L}_u}\left[
t\theta+\bar{t}\bar{\theta}\right],
\end{eqnarray}
where $\theta$ is a Liouville one-form 
\begin{eqnarray}
\theta=\frac{t}{4\pi}\bigl(vdu-udv-d(v\bar{u})\bigr).
\label{Liouville}
\end{eqnarray}
This one-form satisfies $d\theta=\omega$, where $\omega$ is a symplectic
form for the commutation relation (\ref{PB uv}).
In particular, for $k=-i\sigma$, we find
\begin{eqnarray}
S=\frac{k}{2\pi}\left({\rm Vol}(m,\ell)+2\pi^2 i{\rm CS}(m,\ell)\right).
\end{eqnarray}
This result explains the volume conjecture naturally from the
Chern-Simons gauge theory.

The subleading term in the asymptotic expansion (\ref{subleading}) of
the Jones polynomial comes from the one-loop term in the Chern-Simons
gauge theory.  In \cite{Witten1}, it is shown that the one-loop term
coincides with the Reidemeister torsion on $M$.  Therefore, the volume
conjecture is generalized to the subleading order.  The remaining
terms $S_{n}$ ($n=2,\cdots$) are higher loop terms in the perturbative
expansion of the vacuum expectation value of the Wilson loop operator
$\langle W_n(K;q)\rangle$ for the $SU(2)$ Chern-Simons gauge theory on
${\bf S}^3$.

\subsection{AJ conjecture}
In \cite{Garou1,Garou2,Garou-Le}, a conjecture on the constraint for
the Jones polynomial is proposed; this conjecture is called the {\it AJ conjecture}.
The claim of the AJ conjecture is a difference equation
\begin{eqnarray}
&&\hat{A}_K(\hat{m},\hat{\ell};q)J_n(K;q)=0,
\label{AJ-conj}
\\
&&\hat{A}_K(m,\ell;q=1)=A_K(m,\ell).
\end{eqnarray}
The operators $\hat{m}$ and $\hat{\ell}$ satisfy the $q$-Weyl
relation.
\begin{eqnarray}
&& \hat{m}f(n)=q^nf(n), \quad \hat{\ell}f(n)=f(n+1),
\\
&& q\hat{m}\hat{\ell}=\hat{\ell}\hat{m}.
\label{q-Weyl}
\end{eqnarray}
with $q=e^{\frac{2\pi i}{k}}$.

From the canonical commutation relation (\ref{commutation}),
we conclude that the
quantized canonical variables $\hat{m}:=e^{\hat{u}}$ and
$\hat{\ell}:=e^{\hat{v}}$ also satisfy the $q$-Weyl relation.  In
terms of the parametrization of the volume conjecture, the eigenvalue
of $\hat{m}$ is $q^n=e^{u}=m$.  The A-polynomial ${A}_K(e^u,e^{v+\pi
  i})$ is interpreted naturally as the Hamiltonian of the
$(u,v)$-system.  In the classical limit $k\to \infty$ the AJ conjecture 
is trivially satisfied because these canonical variables
satisfy the constraint (\ref{A-polynomial}).  In this respect, the AJ
conjecture is nothing but the quantum Hamiltonian constraint on the
partition function $Z(M_u;q)$.

\subsection{Examples}
The volume conjecture and the AJ conjecture have been checked for many
hyperbolic manifolds. In this section, we shall mainly focus on the
figure-eight knot complement and the SnapPea census manifold $m009$
and summarize the computational results.

\subsubsection{Figure-eight knot complement}
As the first example, we will discuss the figure-eight knot
complement.  The figure-eight knot $4_1$ is neither a torus nor 
a satellite knot.  Therefore, the knot complement admits a hyperbolic 
structure. In fact, it can be decomposed into two ideal tetrahedra.  
The gluing and
completeness conditions are solved uniquely, and all of the face angles
are equal to $\pi/3$. By plugging these face angles into (\ref{ideal
  volume}), we find the volume for the figure-eight knot complement to be
\begin{eqnarray}
{\rm Vol}({\bf S}^3\backslash N(4_1))=6\Lambda(\pi/3)
=2D(e^{\pi i/3})=2,0298832\cdots.
\label{volume4_1}
\end{eqnarray}
\begin{figure}[h]
\hspace*{2cm}
\includegraphics[width=12cm,clip]{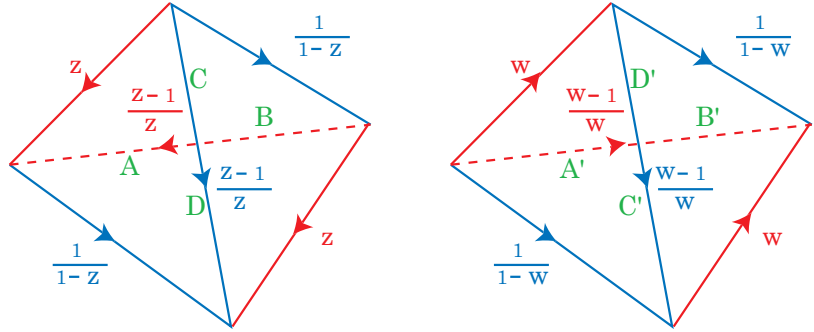}
\caption{Figure-eight knot complement as two ideal tetrahedra.}
\end{figure}
The Wirtinger presentation of the knot group for the figure-eight knot
is
\begin{eqnarray}
&&\pi_1(4_1)=\bigl\langle a,b
\bigl|(ab^{-1}a^{-1}b)a=b(ab^{-1}a^{-1}b)\bigr\rangle,\\
&&
\mu:=a,\quad\lambda:=ab^{-1}aba^{-2}bab^{-1}a^{-1},
\end{eqnarray}
and the A-polynomial is computed as
\begin{eqnarray}
A_{4_1}(m,\ell)=\ell+\ell^{-1}-m^2+m+2+m^{-1}-m^{-2}.
\label{A4_1}
\end{eqnarray}
This A-polynomial is also obtained from the gluing conditions
\cite{Dinesh}.

The logarithmic Mahler measure is given by the period integral
\cite{Boyd,BDR}
\begin{eqnarray}
m({A}_{4_1})=\frac{3\sqrt{3}}{8\pi}L(\chi_{-3},2),
\end{eqnarray}
where $L(\chi_{-f},s)$ is an L-function with Dirichlet character modulo
$f$.
\begin{eqnarray}
L(\chi,s)=\sum_{n=1}^{\infty}\frac{\chi_{-f}(n)}{n^s}
=\prod_{p:{\rm prime}}(1-\chi_{-f}(p)p^{-s})^{-1}.
\end{eqnarray}
In the case $f=3$, the Dirichlet character is
\begin{eqnarray}
\chi_{-3}(n)=0,1,-1, \quad {\rm if}
\quad n\equiv 0,1,2 \quad
 {\rm mod}\;3.
\end{eqnarray}
The numerical value of $\pi m({A}_{4_1})$ coincides with the sum of
dilogarithm functions (\ref{volume4_1}) up to a non-trivial order.  This
arithmetic expression is consistent with the Humbert's volume formula
\cite{Humbert} for Bianchi manifold ${\bf H}^3/\Gamma_f$
\cite{Bianchi} with $f=3$.

Since the figure-eight knot is isomorphic to its mirror image, the
Chern-Simons invariant for ${\bf S}^3\backslash 4_1$ vanishes
\begin{eqnarray}
{\rm CS}({\bf S}^3\backslash 4_1)=0.
\end{eqnarray}

The colored Jones polynomial for the figure-eight knot 
$J_n(4_1;q)$ is \cite{Habiro,Masbaum} 
\begin{eqnarray}
J_n(4_1;q)=\sum_{i=1}^{n-1}\prod_{j=1}^i
\left(
q^{(n+j)/2}-q^{-(n+j)/2}
\right)\left(
q^{(n-j)/2}-q^{-(n-j)/2}
\right).
\label{jones4_1}
\end{eqnarray}
The volume conjecture can be checked explicitly by evaluating the saddle
point contributions. It is also
clear that the Chern-Simons invariant vanishes because $J_n(4_1;q)$ is
real for $q=e^{2\pi i/n}$.  On the basis of this cyclotomic expansion,
it has been shown that the Jones polynominal (\ref{jones4_1}) satisfies the
AJ-conjecture (\ref{AJ-conj}) via the $q$-Zeilberger algorithm
\cite{Garou1,Garou2,Garou-Le}.

The figure-eight knot complement can also be constructed as a once-punctured torus bundle over a circle.  Such a three-manifold
$M_{\varphi}$ is constructed as
\begin{eqnarray}
M_{\varphi}\simeq F\times [0,1]/\{(0,1)\sim (\varphi(x),1)\},
\end{eqnarray}
where the fiber $F$ is a punctured torus $F={\bf T}^2\backslash
\{0\}$.  If the monodromy $\varphi\in PSL(2;\mathbb{Z})$ has two
distinct eigenvalues, $M_{\varphi}$ admits a complete hyperbolic
structure. In general, such $\varphi$ is canonically found as
\begin{eqnarray}
&&\varphi=L^{s_1}R^{t_1}L^{s_2}R^{t_2}\cdots L^{s_r}R^{t_r},
\quad s_i, t_i\in \mathbb{Z}_+,
\\
&& L:=\left(
\begin{array}{cc}
1 & 1 \\
0 & 1
\end{array}
\right),
\quad 
R:=\left(
\begin{array}{cc}
1 & 0 \\
1 & 1
\end{array}
\right).
\end{eqnarray}
In fact, the figure-eight knot complement is a once-punctured torus bundle with
the monodromy $\varphi=LR$.

The Reidemeister torsion for a once-punctured torus bundle over a
circle is studied in Porti's work \cite{Porti}.  For $\varphi=LR$, the
Reidemeister torsion is expressed as a function of the holonomy
\begin{eqnarray}
&&T_{LR}(u)=\frac{1}{\sqrt{(x+1/2)(x-2/3)}},
\quad  x:=\cosh u.
\end{eqnarray}
In \cite{GM}, it is checked numerically that the subleading order term
in the asymptotic expansion (\ref{subleading}) of the colored Jones 
polynomial coincides with the Reidemeister torsion for the figure-eight knot complement.

\subsubsection{The SnapPea census manifold $m009$}
The SnapPea census manifold $m009$ \cite{Weeks} has a finite volume
and a Chern-Simons invariant.  This hyperbolic manifold is also
constructed as a once-punctured torus bundle over a circle, now with the 
holonomy $\varphi=L^2R$.  The volume and the Chern-Simons invariant are
computed by using the generalized volume conjecture \cite{HikamiAJ}
\begin{eqnarray}
&& H_{m009}(0)=\frac{\pi^2}{6}+L\left(\frac{1}{x}\right)
-L\left(\frac{1}{x y^2}\right)-L(x^2 y^2), \\
&&{\rm Vol}(m009)=
D\left(\frac{1}{x}\right)
-D\left(\frac{1}{x y^2}\right)-D(x^2 y^2)
=2,66674\cdots,
\\
&&
{\rm CS}(m009)=0,0208333\cdots,
\end{eqnarray}
where $(x,y^2)=\left(\frac{1\pm i\sqrt{7}}{4}, \frac{-1\pm
  i\sqrt{7}}{4}\right)$ and $L(x)$ is Roger's dilogarithm function
$L(x)=Li_2(x)+\frac{1}{2}\log x\cdot \log(1-x)$.

The fundamental group for $m009$ is
\begin{eqnarray}
\pi_1(m009)=\langle\alpha,\beta,\mu\bigl|
\mu\alpha\mu^{-1}=\alpha\beta,\quad \mu\beta\mu^{-1}=\beta\alpha\beta\alpha\beta
\rangle.
\end{eqnarray}
From this relation, we find the A-polynomial for $m009$ 
\begin{eqnarray}
A_{m009}(m,\ell)=1+(-1+2m+2m^2-m^3)\ell+m^3\ell^2.
\label{Am009}
\end{eqnarray}
There are two solutions $\ell=\ell_{\pm}(m)$ for 
${A}_{m009}(m,\ell)=0$.
\begin{eqnarray}
\ell_{\pm}(m)
=\frac{1-2m-2m^2+m^3\pm (m-1)\sqrt{1-2m-5m^2-2m^3+m^4}}{2m^3}.
\end{eqnarray}
By integrating the Liouville one-form $\theta$ numerially,
we obtain the volume and the Chern-Simons invariant.
\begin{eqnarray}
&& {\rm Vol}(m009)=\frac{1}{2}\left(
\int_{0}^{\pi}d\theta\; \log |\ell_+(e^{i\theta})|
+\int_{0}^{\infty}du \; {\rm arg} (\ell_+(e^{u}))\right),
\\
&& {\rm CS}(m009)=-\frac{1}{4\pi^2}\left(
\int_{0}^{\pi}d\theta\; {\rm arg} (\ell_+(e^{i\theta}))
+\int_{0}^{\infty}du \; \log |\ell_+(e^{u})|\right),
\end{eqnarray}
where the integration path ${\cal L}$ is an average of ${\cal L}_{1}$
and ${\cal L}_{2}$.\footnote{ In terms of Jensen's formula, the volume
is obtained by the integration along ${\cal L}_{1}$ only.  In order
to recover the Chern-Simons invarint, we should take an average of
the integration along ${\cal L}_{1}$ and ${\cal L}_{2}$.  } In
particular, for the volume part, the logarithmic Mahler measure for
${A}_{009}$ is computed exactly as \cite{BDR}
\begin{eqnarray}
m({A}_{m009})=\frac{7\sqrt{7}}{8\pi}L(\chi_{-7},2).
\end{eqnarray}
This result is also consistent with Humbert's formula.
\begin{figure}[h]
\hspace*{1cm}
\includegraphics[width=15cm,clip]{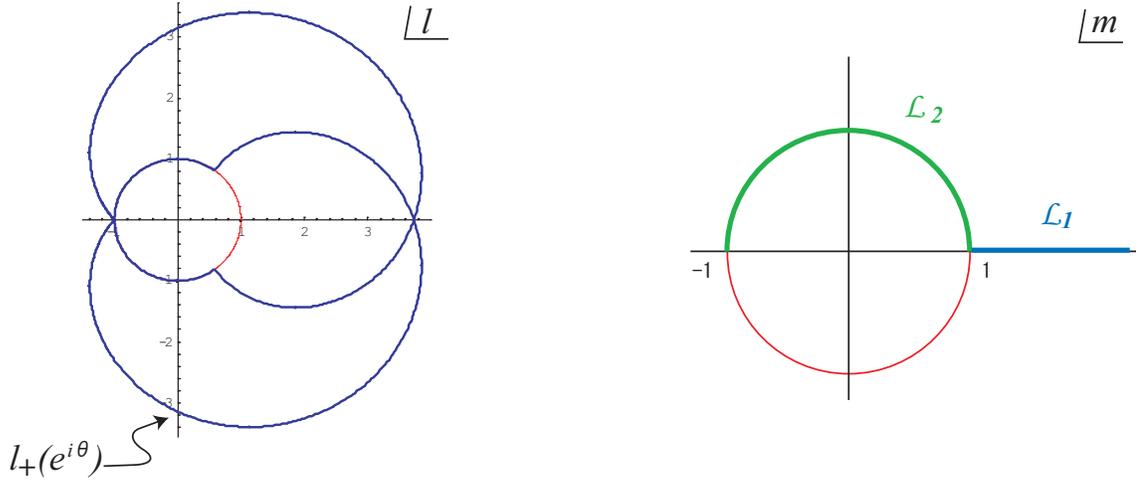}
\caption{Profile of $\ell_+(e^{i\theta})$ and the integration path.}
\end{figure}

The Reidemeister torsion for a once-punctured torus bundle over a circle 
with a holonomy $\varphi=L^2R$ is\footnote{
In \cite{Porti}, the formula for the Reidemeister torsion is given 
as follows:
\begin{eqnarray}
&& I_{\mu}=y_0,\quad I_{\mu\alpha}=y_1,\quad I_{\mu\beta}=y_2,
\quad I_{\mu\alpha\beta}=y_3,\\
&&y_0=y_2, \quad y_1=y_3,\quad 2y_0=x_1y_1, \quad y_0=\frac{1}{2}x_1y_1, \quad (x_1^2-2)y_1^2-4(x_1^2-1)=0, \\
&&T_{L^2R}=\frac{1}{\frac{1}{2}y_1^2-x_1^2}, \quad y_0=2\cosh u/2,
\end{eqnarray}
where $I_{\gamma}:={\rm tr}\rho(\gamma)$, $\gamma\in \pi_1(M)$.
By eliminating parameters $x_1,y_0,y_1,y_2$, and $y_3$, we obtain the formula
(\ref{Reidemeister L2R}).
}
\begin{eqnarray}
T_{L^2R}(u)=\frac{1}{\sqrt{4x^2-4x-7}},\quad x=\cosh u.
\label{Reidemeister L2R}
\end{eqnarray}

\section{Open topological string and Jones polynomial}
\subsection{Disk instanton in topological string}
In the topological string theory, the A-model partition function
$Z_{\rm closed}$ is equivalent to the generating function of the
number of holomorphic maps from the world-sheet Riemann surface 
to a Calabi-Yau threefold $X$.  By introducing D-branes that wrap around a
special Lagrangian submanifold in $X$, we ensure that the open strings end on
them. The partition function for this open string sector is given by
the holomorphic maps for the world-sheet with boundaries ending on 
the Lagrangian submanifold in $X$.  For example, the free energy for the
disk instanton  
is the generating function of the number of holomorphic disks ending 
on the special Lagrangian submanifold \cite{OV,AV}.
\begin{eqnarray}
F_d(X)
&=&\sum_{\beta\in
 H_2(X)}\sum_{m\in {\bf Z}}N_{\beta,m}^{0,1}e^{-\int_{\beta}\omega-mu}
\nonumber \\
&=&\sum_{\beta\in H_2(X)}\sum_{m\in{\bf Z}}\sum_{d>0}n_{\beta,m}^{0,1}
\frac{1}{d^2}e^{-d\int_{\beta}\omega-dmu}. 
\end{eqnarray}
The open Gromov-Witten invariants $N_{\beta,m}^{0,1}$ are defined 
for a relative homology class $(m,\beta)\in H_2(X,L)$,
and they are related to the Ooguri-Vafa invariants 
$n_{\beta,m}^{0,1}\in \mathbb{Z}$ by a resummation.
\begin{figure}[h]
\hspace*{3cm}
\includegraphics[width=8cm,clip]{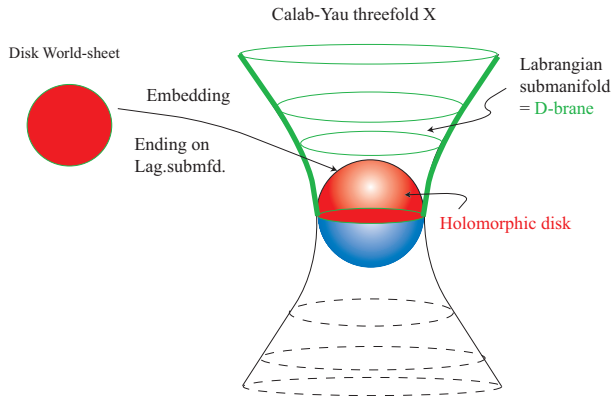}
\caption{Holomorphic disk ending on D-brane.}
\end{figure}

The free energy for the topological string theory is computed 
using the {\it mirror symmetry}.
Mirror symmetry relates the original Calabi-Yau threefold $X$ to
the mirror Calabi-Yau threefold $X^{\vee}$ by exchanging their
cohomology classes $H^{2,1}$ and $H^{1,1}$.  The general form of the
defining equation of the mirror geometry for the non-compact toric
Calabi-Yau threefold is
\begin{eqnarray}
X^{\vee}=\bigl\{
(x,y,z,w)\in(\mathbb{C}^{*})^2\times \mathbb{C}^2\bigl|
zw=H(x,y)
\bigr\}.
\end{eqnarray}
In terms of these coordinates, the holomorphic three-form on $X^{\vee}$
is given by
\begin{eqnarray}
\Omega=\frac{dz\wedge dx\wedge dy}{z}.
\end{eqnarray}
The B-brane in this geometry wraps around the holomorphic curve
${\cal C}$ in $X^{\vee}$.
\begin{eqnarray}
{\cal C}=\bigl\{
(u,v,z,w)\in\mathbb{C}^4\bigl|
H(e^{-u},e^{-v})=0,\quad w=0
\bigr\}.
\end{eqnarray}
Here, we assume that the parameters $x$ and $y$ are flat coordinates
\cite{AKV}.  Geometrically, the variable $u:=\log x$ is the area of a
holomorphic disk where the corresponding B-brane is inserted on an
appropriate patch with some framing.

By reducing the holomorphic Chern-Simons action \cite{Witten4} 
\begin{eqnarray}
S_{\rm hCS}=\frac{1}{g_s}\int_{X^{\vee}}\Omega\wedge {\rm tr}\left(
\bar{A}\wedge \bar{\partial}\bar{A}
+\frac{2}{3}\bar{A}\wedge\bar{A}\wedge\bar{A}
\right),
\end{eqnarray}
on the curve ${\cal C}$, we find the effective action for the
non-compact B-brane as the Abel-Jacobi map
\begin{eqnarray}
S_{\rm hCS}=\frac{1}{g_s}\int_{u_*}^{u} du\;v(u),
\label{rhCS}
\end{eqnarray}
where $H(e^{u},e^{v})=0$ is satisfied.  This relation is analogous to
the Neumann-Zagier function $H(u)$, if we choose
$H(x,y)=A_K(x,y)$.\footnote{ There is a total derivative term
  $d(v\bar{u})$ in the Liouville one-form (\ref{Liouville}). This term
  can be introduced, if we choose a K\"ahler polarization
  \cite{HMurakami}.  }

Concretely, the character variety for the figure-eight knot is
realized in the mirror geometry of the $A_3$-fibration over ${\bf
  P}^1$. The toric vectors are \cite{CKYZ}
\begin{eqnarray}
\ell^{(1)}=(0,0,1,-2,1,0,0),\quad \ell^{(2)}=(0,0,0,1,-2,1,0),
\quad \ell^{(3)}=(0,0,0,0,1,-2,1),
\nonumber 
\end{eqnarray}
and
\begin{eqnarray}
 \ell^{(4)}=(1,1,0,0,-2,0,0).
\end{eqnarray}
From the mirror map, we find the defining equation for the mirror
Calabi-Yau threefold $X^{\vee}$. By adjusting the complex structure
moduli appropriately, we can realize $H(x,y)={A}_{4_1}(x,y)$ from this
geometry.
\begin{figure}[h]
\hspace*{2cm}
\includegraphics[width=12cm,clip]{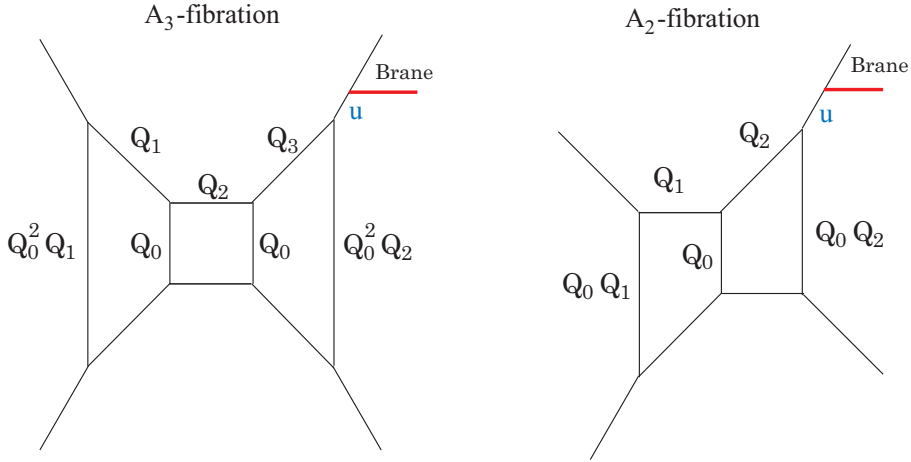}
\caption{Holomorphic disk ending on D-brane.}
\end{figure}
In the same manner, the character variety for $m009$ is realized in
the mirror geometry of the $A_2$-fibration over ${\bf P}^1$, whose
toric vectors are
\begin{eqnarray}
\ell^{(1)}=(0,0,1,0,-2,1),\quad\ell^{(2)}=(0,0,0,1,1,-2),
\quad {\rm and}
\quad\ell^{(3)}=(1,1,0,0,-1,-1).
\end{eqnarray}

\subsection{Kodaira-Spencer theory and ${\cal D}$-module structure}
The quantum nature of the non-compact B-brane is understood 
from the Kodaira-Spencer gravity on the Riemann surface $\Sigma$
\cite{BCOV,ADKMV,DVKS}.
\begin{eqnarray}
\Sigma:=\bigl\{
(u,v)\in\mathbb{C}^{2}\bigl|
H(e^{-u},e^{-v})=0
\bigr\}.
\end{eqnarray}
The reduction of the holomorphic three-form $\Omega$ on $\Sigma$ gives
rise to the Liouville one-form $\theta$,
\begin{eqnarray}
\theta=vdu.
\end{eqnarray}
The Kodaira-Spencer field causes the deformation of the
complex structure of the mirror Calabi-Yau threefold $X^{\vee}$, and
one of the coordinates which is dual to the disk moduli $u$ is related as
\begin{eqnarray}
v=\partial \phi.
\label{phi}
\end{eqnarray}
In the classical limit $g_s\to 0$, the expectation value of the
Kodaira-Spencer field satisfies
\begin{eqnarray}
&& \langle \phi(e^{-u})\rangle=:\phi_{\rm cl}(e^{-u}), \quad 
H(e^{-u},e^{-\partial \phi_{\rm cl}})=0, \\
&& \phi_{\rm cl}(e^{-u})=\int^u_{u_{*}} du\;v(u).
\end{eqnarray}
Because $(u,v)$ can be identified with the canonical coordinate and momentum,
the semi-classical value of the disk free energy $F_d$ can be given by
\begin{eqnarray}
F_d=\frac{1}{g_s}\phi_{\rm cl}.
\end{eqnarray}

This semi-classical analysis implies that the D-brane partition
function $Z_D(u):=Z_{\rm closed}\cdot Z_{\rm open}(u)$ is expressed as
\begin{eqnarray}
Z_{D}(u)=\bigl\langle
e^{\phi(e^{-u})/g_s}
\bigr\rangle_{\Sigma}.
\end{eqnarray}
In the quantum theory of the chiral boson $\phi(e^{-u})$ on 
$\Sigma$, the operator $e^{\phi(e^{-u})/g_s}$ is identical 
to a fermionic field
$\psi(e^{-u})$ via the bosonization. 
Thus, we find that the D-brane partition function 
can be seen as a Baker-Akhiezer function \cite{ADKMV,Vertex}. 

This relation (\ref{phi}) can be naturally identified with the Hamilton-Jacobi
equation with the canonical coordinate $u$, momentum $v$, and 
Hamiltonian $H(e^{-u},e^{-v})$.  The canonical pair $(u,v)$ is
quantized with respect to the reduced action (\ref{rhCS}).
\begin{eqnarray}
[\hat{u},\hat{v}]=g_s.
\label{comm}
\end{eqnarray}
The coordiantes $\hat{x}:=e^{-\hat{u}}=e^{-u-g_s/2}$ and
$\hat{y}:=e^{-\hat{v}}=e^{g_s\partial_u}$ satisfy the $q$-Weyl
relation
\begin{eqnarray}
\hat{x}\cdot \hat{y}=q\;\hat{y}\cdot \hat{x}, \quad q=e^{g_s}. 
\end{eqnarray}
In the crystal melting description \cite{ORV} of the topological 
string theory,
the non-compact D-brane insertion is realized by introducing the small
defect $\Psi_D(e^u):=\Gamma_-^{-1}(e^u)\Gamma_+(e^u)$ at
$u:=g_s(N_0+1/2)$, where $\Gamma_{\pm}(z):=e^{\pm \phi(z)}$
\cite{Saulina-Vafa}.  In this set-up, we find the discretized
$q$-Weyl relation (\ref{q-Weyl}).

In the topological B-model, there exists a global diffeomorphism that
preserves the choice of $\Omega$ of the Calabi-Yau threefold
$X^{\vee}$. 
By reducing this symmetry onto $\Sigma$,
we find the action of the $W_{1+\infty}$ algebra, which is generated by
\begin{eqnarray}
W^{n+1}\sim \frac{1}{n+1}(\partial \phi)^{n+1},\quad  (n=1,\cdots,\infty).
\end{eqnarray}
However, this $W_{1+\infty}$ symmetry is broken in the presence of the
curve $H(e^{-u},e^{-v})=0$.  From the operator $\hat{\cal O}$ in the
original symmetry, the operators $\hat{\cal O}_{\rm unbroken}$ for the
unbroken symmetries are constructed as
\begin{eqnarray}
\hat{\cal O}_{\rm unbroken}
=\hat{H}(e^{-\hat{u}},e^{-\partial\hat{\phi}};q)\cdot \hat{\cal O}, 
\end{eqnarray}
where $\hat{H}(x,y;q=1)=H(x,y)$.\footnote{There exists a normal ordering
ambiguity for the definition of the quantum Hamiltonian operator.}
From the commutation relation (\ref{comm}), 
we find the constraint equation for the D-brane partition function 
\begin{eqnarray}
\hat{H}(e^{-u-g_s/2},e^{g_s\partial_u};q)Z_D(u;q)=0.
\end{eqnarray}
In the canonical picture, this constraint equation is naturally
understood as the the Schr\"odinger equation.\footnote{ In terms of
  the fermion one-point function, this is nothing but the Lax
  equation.}   We can check this ${\cal D}$-module structure of the
D-brane partition function explicitly, if $\Sigma$ is a genus zero curve
\cite{Kashani,Hyun-Lee}, although it is conjectural for higher genus
case \cite{DHSV,DHS,Bethe1}. 
Compared to the three-dimensional Chern-Simons gauge theory results, this constraint equation is
naturally
identified with the AJ conjecture (\ref{AJ-conj}).

\subsection{Correspondence with three dimensional theory}
Here we find some similar structures in the three
dimensional Chern-Simons gauge theory and the topological open string theory.
The first fact is the correspondence of the free energies 
in the asymptotic limit of the both theories.
In the asymptotic limit,
the partition function of the three dimensional Chern-Simons gauge
theory is given by the integral of the Liouville one-form on the phase
space which is determined by the A-polynomial. On the other hand, the free
energy for the disk instantons in the topological string is given by
the Abel-Jacobi map on the Riemann surface $\Sigma$ inside the mirror
Calabi-Yau geometry.
\begin{eqnarray}
F_{\rm cl}\sim\int_{\gamma\subset \Sigma} v du.
\end{eqnarray}

Although we find the similar integral forms in both theories,
the path $\gamma$ in the integration is different.
In WKB analysis, the leading term of the free energy should be evaluated at
the saddle point of the partition function. 
Then, the integration path $\gamma$ will change for the value of 
the coupling constant.
For the topological string, the integration path
is choosen as the real loci of $(u,v)$ cooridinates, because the
string coupling $g_s$ is real$\mathchar`-$valued parameter 
\cite{AV,Saulina-Vafa}.
On the other hand, the $SL(2,\mathbb{C})$ Chern-Simons gauge theory is
expanded with respect to the complex coupling $\frac{4\pi i}{t}$.
In the asymptotic limit, the volume term is given by the special value
of the Ronkin function.\footnote{It is not obvious whether one can
always find the appropriate path which gives rise to the Chern-Simons 
invariant correctly. At least for SnapPea census manifold $m009$, 
we can find the natural path ${\cal L}_1$ and ${\cal L}_2$.}
Such difference of the coupling constants leads to the difference of the
choice of the integration paths. Therefore, we expect that the leading
terms in the WKB expansion of both theories will correspond under the
appropriate analytic continuation.

The endpoint of the path is determined by the critical value $u=u_0$
of the Neumann-Zagier function in the volume conjecture.
In the topological open string theory,
the similar extremization is considered in
\cite{Jockers-Soroush}$\mathchar`-$\cite{AHMM}.
In this context, the extremization of the superpotential freezes
the open string moduli $u$,
and the critical value of the superpotential gives rise to
the generating function for the real BPS numbers
\cite{Walcher1}$\mathchar`-$\cite{Knapp-Scheidegger}.
Thus, the minimization of the free energy are meaningful in both theories.

Furthermore, we also find the correspondence of the quantum Hamiltonian
structure in both theories. The Hamiltonian on the phase space of the three dimensional Chern-Simons gauge theory is given by the
A-polynomial $A_K$. On the other hand, the Hamiltonian for the Kodaira-Spencer
theory is given by the non-trivial polynomial $H$ in the defining equation of 
the mirror Calabi-Yau geometry. Both of the Hamiltonian structures lead
to the ${\cal D}$-module structure, in particular, 
the non-perturbative Hamiltonian constraint 
on the full partition function.
Up to the normal ordering ambiguities, we expect that
the correspondence will hold beyond the WKB analysis under the appropriate
analytic continuation.

From these facts, we find the following correspondences between 
the three-dimensional Chern-Simons gauge theory and the topological 
open string theory.
\begin{center}
\begin{tabular}{c|c}
{\bf 3D Chern-Simons} & {\bf  Topological Open String} \\ \hline
$u=2\pi i \left(\frac{n}{k}-1\right)$: Meridian holonomy  &
 $u$: Area of holomorphic disk \\
$q=e^{2\pi i/k}$ & $q=e^{g_s}$ \\
Vol+$i$CS &
Disk free energy ${\cal F}_d$ \\
AJ conjecture & Schr\"odinger equation \\
${A}_K(\hat{m},\hat{\ell};q)J_n(K;q)=0$ & 
 $\hat{H}(e^{-u-g_s/2},e^{g_s\partial_u};q)Z_{\rm open}(u;q)=0$ \\
$\hat{m}\hat{\ell}=q\hat{\ell}\hat{m}$ & 
$e^{-u-g_s/2}e^{g_s\partial_u}=qe^{g_s\partial_u}e^{-u-g_s/2}$ \\
\end{tabular}
\end{center}
In the basis of these correspondences, we can read off a relation,
\begin{eqnarray}
J_{n}(K;q=e^{2\pi\sqrt{-1}/k})
\simeq
Z_{\rm open}(e^{u};q),
\label{CONJECTURE}
\end{eqnarray}
where $Z_{\rm open}$ is an open string partition function on the mirror
Calabi-Yau threefold $X^{\vee}$
\begin{eqnarray}
X^{\vee}=\{(x,y,z,w)\in(\mathbb{C}^{*})^2\times \mathbb{C}^2
|zw={A}_K(x,y)\}.
\end{eqnarray}

In the basis of the correspondence (\ref{CONJECTURE}), the subleading terms
in the WKB expansion of the Chern-Simons gauge theory and the topological
open string theory should also coincide.  In the following, we will
check this correspondence in the subleading order.

\section{Computation via Chern-Simons matrix model}
In the topological string theory, the subleading contributions 
come from the world-sheet instanton with the
annulus topology.  Now, we discuss the correspondence between the
Reidemeister torsion of the hyperbolic manifold
and the annulus free energy in the topological string
theory for the figure-eight knot complement and the SnapPea census
manifold $m009$ via the Chern-Simons matrix model.

\begin{figure}[h]
\hspace*{2cm}
\includegraphics[width=12cm,clip]{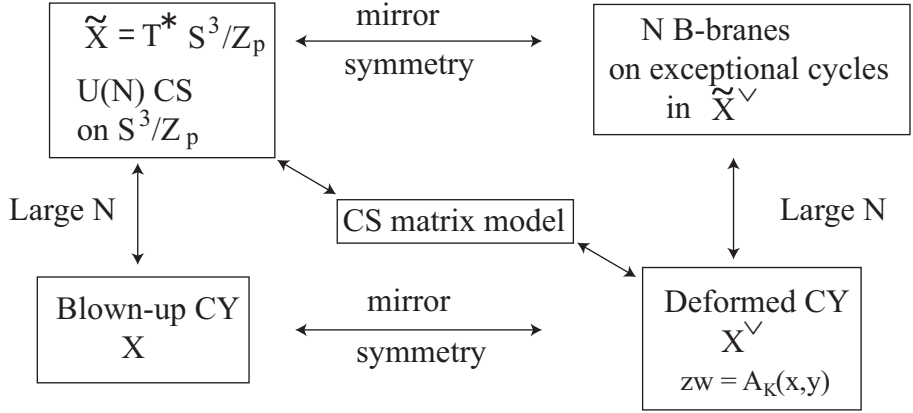}
\caption{Web of dualities in \cite{AKMV}.}
\end{figure}
Through the web of dualities \cite{AKMV}, the topological string theory 
is mapped to $U(N)$ Chern-Simons gauge theory on a compact Lagrangian
submanifold $L$ in a non-compact Calabi-Yau threefold $T^*L$.\footnote{In
the recent paper \cite{BGST}, it is pointed out that the web of the 
dualities is not clear for $L=L(p,q)$ $q\ne 1$. In this paper, we use
only the $L(p,1)$ matrix models to check our proposal.} 
The $U(N)$ Chern-Simons partition function for the Lens space $L(p,q)$
with a fixed flat connection ${\bf n}\in \mathbb{Z}_p^{\otimes N}$ is
given by \cite{BGST,Hansen-Takata,GSST}
\begin{eqnarray}
&&\hspace*{-0.8cm}
Z(L(p,q);{\bf n})=C_N(p,q;g_s)e^{-\frac{4\pi^2q}{g_s^2p}{\mathbb{\bf n}^2}}
\sum_{w,w^{\prime}\in S_N}\epsilon(w)\exp\left[
\frac{g_s^2}{2p}w(\rho)\cdot \rho+\frac{2\pi i}{p}w^{\prime}({\bf
n})\cdot (q\rho+w(\rho)) 
\right],
\nonumber \\
&&
\end{eqnarray}
where $\rho$ is the Weyl vector $\rho:=\sum_{\alpha>0}\alpha$ for $SU(N)$ group and $g_s$ is related
to the level $k$ of the $U(N)$ Chern-Simons gauge theory
by $g_s^2:=\frac{4\pi i}{k+N}$.
The factor $C_N(p,q;g_s)$ does not depend on the choice of the flat
connection ${\bf n}$.

Here we neglect the factor $C_N(p,q;g_s)$, and rewrite this expression by
the $N$-dimensional integral form. We obtain the matrix model-like 
integral form by using the Weyl formula \cite{BGST,Okuda}.
In particular, for the Lens space $L(p,1)$, 
we obtain \cite{Marino,AKMV,remodel}.
\begin{eqnarray}
Z(L(p,1))=\int\prod_{i=1}^{N}du_i \prod_{j<k}2\sinh^2\frac{u_j-u_k}{2}
\exp\left[-\frac{p}{g_s}\sum_{i=1}^N \left(\frac{u_i^2}{2}
-\frac{2\pi i}{p}n_i u_i \right)\right].
\end{eqnarray} 
 This model is called the {\it  Chern-Simons matrix model}. 
The details of the large $N$ analysis of
the Chern-Simons matrix model is summarized in Appendix.  The
spectral curve $\Sigma_p$ for this matrix model is \cite{HY,HOY,BGST}
\begin{eqnarray}
\Sigma_p:=\bigl\{(u,v)\in \mathbb{C}^2\bigl|
(e^{v}-1)(e^{pz}e^v-1)+e^S-1+e^v\sum_{n=1}^{p-1}d_ne^{nz}=0 \bigr\},
\label{spectral0}
\end{eqnarray}
where the coefficients $d_n$ are determined uniquely by specifying
the number of the eigenvalues $u_i$ on each cut in $\Sigma_p$, and
$S:=g_sN$.  By chaising the web of dualities, one finds that the original
B-model geometry $X^{\vee}$ is the $\mathbb{C}^2$ fibration over
$\Sigma$.
\begin{eqnarray}
X^{\vee}=\bigl\{(u,v,z,w)\in \mathbb{C}^4\bigl|
zw=
(e^{v}-1)(e^{pz}e^v-1)+e^S-1+e^v\sum_{n=1}^{p-1}d_ne^{nz} \bigl\}.
\label{CSCY}
\end{eqnarray}
In fact, the partition function $Z(L(p,1))$
coincides with that of the topological B-model on $X^{\vee}$ \cite{AKMV}.  
By adjusting the eigenvalue distributions, we find that
the spectral curves $\Sigma_p$ for $p=4$
and $p=3$ coincide with the character variety for figure eight knot
complement (\ref{A4_1}) and the SnapPea census manifold $m009$
(\ref{Am009}), respectively.

If the non-compact D-brane is introduced, the open string modes that
connect $L(p,1)$ and the non-compact D-brane appear. The physical states
of the topological open string come from the complex scalar fields
$\varphi$ and $\bar{\varphi}$.  These scalar fields lead to the
following effective action \cite{OV,Hyun-Lee}.
\begin{eqnarray}
\int {\cal D}\bar{\varphi}{\cal D}\varphi \exp\left[
-\bar{\varphi}(V\otimes \mathbb{I}_{N\times N}
-\mathbb{I}_{M\times M}\otimes U
)\varphi
\right],
\end{eqnarray}
where $M$ is the number of non-compact D-branes, and $U$ and $V$ are
matrices that result from the path-ordered exponentials of the gauge
fields on the compact and non-compact branes, respectively.  By evaluating
the path integrals for scalar fields, we obtain the D-brane partition
function
\begin{eqnarray}
Z_D(u)=\langle \det(\mathbb{I}_N-e^{-u}U^{-1})^{-1}\rangle.
\end{eqnarray}
In the case of an anti-brane, the partition function is
\begin{eqnarray}
Z_{\bar{D}}(u)=\langle \det(\mathbb{I}_N-e^{-u}U^{-1})\rangle.
\end{eqnarray}

In the large $N$ limit, the free energy of ${\cal O}(N)$ is simply
computed as
\begin{eqnarray}
F_d(u)=\int_{\infty}^u du\;v(u).
\end{eqnarray}
This result coincides with the disk free energy on $X^{\vee}$
(\ref{CSCY}).  From the facts summarized in section 2, we find
the volume, Chern-Simons invariant, and Neumann-Zagier
function for the hyperbolic three-manifolds from the topological string
theory by selecting the integration path ${\cal L}_u$ appropriately.  
This fact supports our proposal in the case of the asymptotic limit.

The subleading terms can also be computed in the large $N$ analysis of
the Chern-Simons matrix model. The annulus free energy $F_a$ is
\cite{FM}
\begin{eqnarray}
F_a(u)=\frac{1}{2}\int_{e^u}^{\infty} \omega_{{e^{\tilde{u}}}-\tilde{\infty}}
=\frac{1}{2}\log\frac{E(e^{-u},\tilde{\infty})E(\infty,e^{-\tilde{u}})}{E(e^{-u},e^{-\tilde{u}})E(\infty,\tilde{\infty})},
\label{annulus}
\end{eqnarray}
where $e^{-\tilde{u}}$ and $\tilde{\infty}$ represent the points on the
 covering space of the spectral curve $\Sigma_p$.
$\omega_{a-b}$ is the abelian differential of the third kind on
$\Sigma_p$, which has zero A-periods and simple poles at $a$ and $b$
with residues $+1$ and $-1$, respectively.
$E(a,b)\sqrt{dz(a)}\sqrt{dz(b)}$ is the prime form on $\Sigma_p$.  This
integral can also be expressed as $F_a\sim\int B(x_1,x_2)$, where
$B(x_1,x_2)$ is the Bergman kernel \cite{Fey,Tata}.  This result is
consistent with the annulus amplitude of the topological string theory
\cite{Marino2,DVKS,remodel,Eynard-Orantin}.  By using the matrix model formula
(\ref{annulus}), we will analyze the annulus free energy for $H={A}_K$
case.
\begin{figure}[h]
\hspace*{3cm}
\includegraphics[width=8cm,clip]{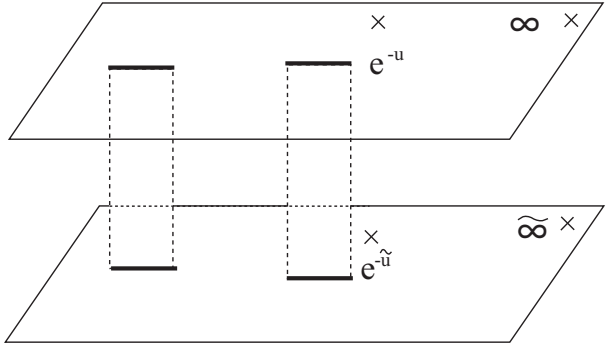}
\caption{Double covering of the spectral curve.}
\end{figure}

\subsection{Reidemeister torsion for figure-eight knot complement}
By neglecting the overall factor $m^{-2}\ell^{-1}$, we can express 
the A-polynomial ${A}_K$ for the figure-eight knot complement as
\begin{eqnarray}
{A}_{4_1}(m,\ell)=m^4\ell^2+(-m^4+m^3+2m^2+m-1)\ell+1.
\label{reduced4_1}
\end{eqnarray}
The character variety for $X_{4_1}$ has a singular point $m=1$.  In the
matrix model analysis, if there exists a classical critical point
around which no eigenvalue is filled, then the cut is not created on
the complex plane and a singular point appears in the spectral curve
\cite{BIPZ}.  The spectral curve (\ref{spectral0}) for $p=4$ is
\begin{eqnarray}
\Sigma_4=\{(x,y)\in (\mathbb{C}^{*})^2|
x^4 y^2+(-x^4+d_3x^3+d_2x^2+d_1x-1)y+e^S=0
\}.
\end{eqnarray}
By comparing $\Sigma_4$ with the curve $A_{4_1}(x,y)=0$, we have to
  choose $S=0$ to obtain the A-polynomial (\ref{reduced4_1}).\footnote{ In the ordinary matrix model analysis, one cannot
  choose $N=0$ because $N_I<0$ is not allowed.  However in order to describe the
  topological string theory, we should treat the Chern-Simons matrix
  model as the holomorphic matrix model \cite{DV1,DV2,DV3}.  Therefore,
  we can choose $N=0$ by allowing $N_I<0$ configuration by the
  analytic continuation.  } 
Therefore, we should compute the free energy from the
effective curve $\tilde{\Sigma}_{4_1}$, which is obtained by
factoring the term $(x^2-1)$ in the discriminant of (\ref{reduced4_1}) \cite{CV}.
\begin{eqnarray}
\tilde{\Sigma}_{4_1}:=\left\{(x,y)\in (\mathbb{C}^{*})^2
\Bigl|
y^2=x^2-2x-1-2x^{-1}+x^{-2}
\right\}.
\label{reduced 4_1}
\end{eqnarray}
For this curve, the annulus formula (\ref{annulus}) is explicitly
written as
\begin{eqnarray}
F_a(m=e^{-u})=\frac{1}{2}\int_{m}^{\infty} \left[
\frac{dx}{x-m}
\left(1-
\sqrt{\frac{m^2-2m-1-2m^{-1}+m^{-2}}{x^2-2x-1-2x^{-1}+x^{-2}}}
\right)+\omega_{\rm hol}\right].
\end{eqnarray}
$\omega_{\rm hol}$ is the holomorphic one-form 
$\omega_{\rm hol}=\sum_{i=1}^{p-1}a_i\frac{x^{i-1}dx}{y}$, where 
the coefficinets $a_i$ are determined uniquely by the zero A-period
conditions $\oint_{A_i}\omega_{e^{-\tilde{u}}-\tilde{\infty}}=0$.

Here, we reconsider the integration path. In order to recover the
volume conjecture, we should consider a particular analytic
continuation. In the disk case, if the integration path is changed to
${{\cal L}}_u$, we can recover the Neumann-Zagier function.
Furthermore, the disk free energy is not changed, if one changes the
integration path to an average of the ${{\cal L}}_u$ and ${{\cal
    L}}_{-u}$, because the A-polynomial is reciprocal. 
Consequently,
the annulus free energy becomes
\begin{eqnarray}
&& F_a(u)=\frac{1}{4}\left[
\int_m^{x_{-}}+\int_{m^{-1}}^{x_{-}^{-1}}\right]
\omega_{\tilde{m}-\tilde{\infty}}
\nonumber \\
&& =\frac{1}{4}\int_m^{x_{-}}dx
\left(\frac{1}{x-m}+\frac{1}{x-m^{-1}}-\frac{1}{x}
\right)\left(
1-\sqrt{\frac{m^2-2m-1-2m^{-1}+m^{-2}}{x^2-2x-1-2x^{-1}+x^{-2}}}
\right)+\tilde{\omega}_{\rm hol},
\nonumber \\
&&
\end{eqnarray} 
where $x_{\pm}$ are the end points of the cuts in the effective curve
(\ref{reduced 4_1}).  By introducing new variables
$w:=\frac{x+x^{-1}}{2}$ and $a:=\frac{m+m^{-1}}{2}$, we can rewrite
the above expression as
\begin{eqnarray}
F_a(u)&=&\frac{1}{2}\int_{a}^{w_-}\frac{dw}{w-a}\left(
1-\sqrt{\frac{(2a-3)(2a+1)}{(2w-3)(2w+1)}}
\right)+\omega_{\rm hol}
\nonumber \\
&=&\frac{1}{2}\int_{a}^{\infty}\tilde{\omega}_{\tilde{a}-\tilde{\infty}}.
\end{eqnarray}
where $w_-=\frac{x_-+x_-^{-1}}{2}=-1/2$ and
$\tilde{\omega}_{\tilde{a}-\tilde{\infty}}$ is an abelian differential
of the third kind on the Riemann surface $\Sigma^{\prime}_{4_1}$.
\begin{eqnarray}
\Sigma^{\prime}_{4_1}:=\left\{
(w,y)\in \mathbb{C}^2|y^2=(2w-3)(2w+1)
\right\}.
\end{eqnarray}
Because $\Sigma^{\prime}_{4_1}$ is the genus zero curve, the prime form
can be computed exactly as
\begin{eqnarray}
&&E(a,\tilde{a})
=\frac{\sqrt{(2a-3)(2a+1)}}{\sqrt{dw}\sqrt{d\tilde{w}}},
\quad 
E(a,\tilde{\infty})
=\frac{\frac{a}{2}-\frac{1}{4}+\frac{1}{2}\sqrt{(2a-3)(2a+1)}}
{\sqrt{dw}\sqrt{d\tilde{w}}},
\\
&& 
E(w_-,\tilde{a})
=\frac{\frac{a}{2}-\frac{1}{4}-\frac{1}{2}\sqrt{(2a-3)(2a+1)}}
{\sqrt{dw}\sqrt{d\tilde{w}}},\quad  
E(w_-,\tilde{\infty})
=\frac{-1/4}{\sqrt{dw}\sqrt{d\tilde{w}}}.
\end{eqnarray}
Thus, the analytically continued annulus free energy becomes
\begin{eqnarray}
F_a(u)=
\frac{1}{2}\log\frac{E(a,\tilde{\infty})E(w_-,\tilde{a})}{E(w_-,\tilde{a})E(w_-,\tilde{\infty})}
=\frac{1}{2}\log\frac{1}{\sqrt{(2a-3)(2a+1)}},
\end{eqnarray} 
where $a=\frac{e^u+e^{-u}}{2}=\cosh u$.
This result coincides with the Reidemeister torsion $T_{LR}(u)$
\begin{eqnarray}
F_a(u)=\frac{1}{2}\log T_{LR}(u).
\end{eqnarray}
Since the Reidemeister torsion is the subleading term in the
asymptotic expansion (\ref{subleading}), this identity supports our
proposal (\ref{CONJECTURE}).

\subsection{Reidemeister torsion for $m009$}
By neglecting a trivial factor, we ensure that the A-polynomial for $m009$ obeys
\begin{eqnarray}
{A}_{m009}(m,\ell)=m^3\ell^2+(-m^3+2m^2+2m-1)\ell+1.
\label{m009_A}
\end{eqnarray}
In contrast, the spectral curve of the Chern-Simons matrix model
for $p=3$ is
\begin{eqnarray}
\Sigma_3=\left\{
(x,y)\in \left(
\mathbb{C}^{*}\right)^2\Bigl|
x^3 y^2+(-x^3+d_2x^2+d_1x-1)y+e^S=0
\right\}.
\end{eqnarray}
After specifying the parameters $d_i$ and $S$ by adjusting the filling
fractions of the matrix model, we obtain the A-polynomial for $m009$.
By factorizing a singular point $m=1$, we find a smooth effective curve
from the discriminant of (\ref{m009_A}).
\begin{eqnarray}
\Sigma_{m009}=\left\{
(x,y)\in \left(
\mathbb{C}^{*}\right)^2\Bigl|
y^2=x^2-2x-5-2x^{-1}+x^{-2}
\right\}.
\end{eqnarray}
On the basis of the formula (\ref{annulus}), the annulus free energy is given
by
\begin{eqnarray}
F_a(u)=\int_m^{\infty}\left[
\frac{dx}{x-m}\left(
1-\sqrt{\frac{m^2-2m-5-2m^{-1}+m^{-2}}{x^2-2x-5-2x^{-1}+x^{-2}}}
\right)+\omega_{\rm hol}\right].
\end{eqnarray}

We perform the change of the integration path as in the figure-eight
knot case.  Therefore, the annulus free energy takes the form
\begin{eqnarray}
F_a(u)=\frac{1}{2}\int_{a}^{w_-}\left[
  \frac{dw}{w-a}\left(1-\sqrt{\frac{4a^2-4a-7}{4w^2-4w-7}}\right)+\tilde{\omega}_{\rm
    hol}\right],
\end{eqnarray}
where $w=\frac{x+x^{-1}}{2}$, $a=\frac{m+m^{-1}}{2}$, and
$w_{\pm}=\frac{1}{2}(1\pm2\sqrt{2})$.  This integral gives the
prime form on the genus zero curve $\Sigma_{m009}^{\prime}$
\begin{eqnarray}
\Sigma_{m009}^{\prime}=\left\{
(w,y)\in \mathbb{C}^2\Bigl|
y^2=4w^2-4w-7
\right\}.
\end{eqnarray}
Finally, the annulus free energy is given by
\begin{eqnarray}
F_a(u)=\frac{1}{2}\log\frac{1}{\sqrt{4a^2-4a-7}}, \quad a=\cosh u.
\end{eqnarray} 
This result is consistent with the expression of the Reidemeister
torsion $T_{L^2R}(u)$ (\ref{Reidemeister L2R}).
\begin{eqnarray}
F_a(u)=\frac{1}{2}\log T_{L^2R}(u).
\end{eqnarray}
This coincidence also supports our proposal (\ref{CONJECTURE}).

\subsubsection{General case}
In the abovementioned two examples, our proposal is checked up to a one-loop
level.  On the basis of the topological string analysis \cite{remodel}, we
expect the following structure for the subleading term in the
asymptotic expansion of the colored Jones polynomial in general.
\begin{eqnarray}
T_K(u)\sim \frac{1}{\sqrt{E(e^{u},e^{\tilde{u}})E(e^{-u},e^{\tilde{u}})}},
\label{Reidemeister conj}
\end{eqnarray}
where the prime forms $E(x,y)$ are defined on the character variety
$X_{K}$.  Although there exist some normarization factors or
ambiguities of the integration path in (\ref{Reidemeister conj}), the
essential contribution may be given by (\ref{Reidemeister conj}).

\section{Conclusions and discussions}
In this paper, we proposed a correspondence between the colored Jones
polynomial for the hyperbolic knots and the partition function of the
topological open string by using the volume conjecture and the AJ conjecture.
These two different theories have a similar Hamiltonian structure in
their effective descriptions.  On the basis of our proposal, the Reidemeister
torsion for the hyperbolic three-manifold should be described as the 
prime form on the character variety. 
We checked our proposal for the
figure-eight knot complement and the SnapPea census manifold $m009$.
Although there exist some subtle points in the determination of the 
analytically continued integrating paths, we could fix them 
appropriately and obtain the correct result.

The WKB expansion of the topological string has been studied in various
ways. In particular, for the B-model, the recursion relation of
the topological expansion is developed remarkably
\cite{Eynard-Orantin,remodel}.
On the basis of our proposal, the colored Jones polynomial will be closely
related to the correlation funtions $\omega_{g,n}$ for the 
character variety $X_K$. 
Recently,
the ${\cal D}$-module structure of the B-model 
has been discussed in more universal way \cite{Bethe1}.
It is shown that the system must obey the Bethe ansatz equation
in order to satisfy the quantum Riemann surface equation.
In contrast, the gluing conditions (\ref{edge}) 
for the face angles of the ideal tetrahedra in the simplicial
decomposition of the hyperbolic three-manfold
can be further rewritten as \cite{NZ}
\begin{eqnarray}
\prod_{I}z_{I}^{\alpha_{IJ}}(1-z_I)^{\beta_{IJ}}=\pm 1,
\end{eqnarray}
where $\alpha_{IJ}$ and $\beta_{IJ}$ are paired symplectically.
As is indicated in \cite{Bethe2}, these relations resemble to the
Bethe ansatz equation and the volume formula corresponds to the formula
for the central charge. 
From these facts the integrable structure of the colored Jones 
polynomial may be closely related with the hyperbolic structure 
of the three manifold.
The free fermion description of the $SL(2;\mathbb{C})$ Chern-Simons 
gauge theory and its ${\cal D}$-module structure will be reported in the future
works \cite{DF}.


We expect that our proposal could be derived from some explicit string
duality.  In the volume conjecture, we take the large $k$ and $n$
limits, which are the level and the rank of the representation of the Wilson
loop along the knot $K$ in the $SU(2)$ Chern-Simons gauge theory.  
In contrast, the $U(n)$ Chern-Simons gauge theory on $T^{*}L(p,1)$
is related with the topological B-model on the mirror Calabi-Yau $zw=A_K(x,y)$
via the web of dualities for two examples mentioned above.
Then, we expect that some duality, like
the {\it level-rank duality}, might be realized as the large $k$ brane
dynamics.  In fact, in the case of prime $p$, the orbifold limit
($S=0$, $d_{\ell}=0$) of the spectral curve of the Chern-Simons matrix
model is nothing but the character variety for the torus knot that is
obtained by the surgery of the Lens space with an unknot \cite{Adams}.
\begin{eqnarray}
A_{T_{2,p}}=(\ell-1)(\ell m^p-1).
\end{eqnarray}
In general, we also expect that the dual geometry
of the mirror Calabi-Yau $zw=A_K(x,y)$ 
may be related with the $T^* S^3$ by a surgery along the knot $K$.
\begin{figure}[h]
\hspace*{3cm}
\includegraphics[width=10cm,clip]{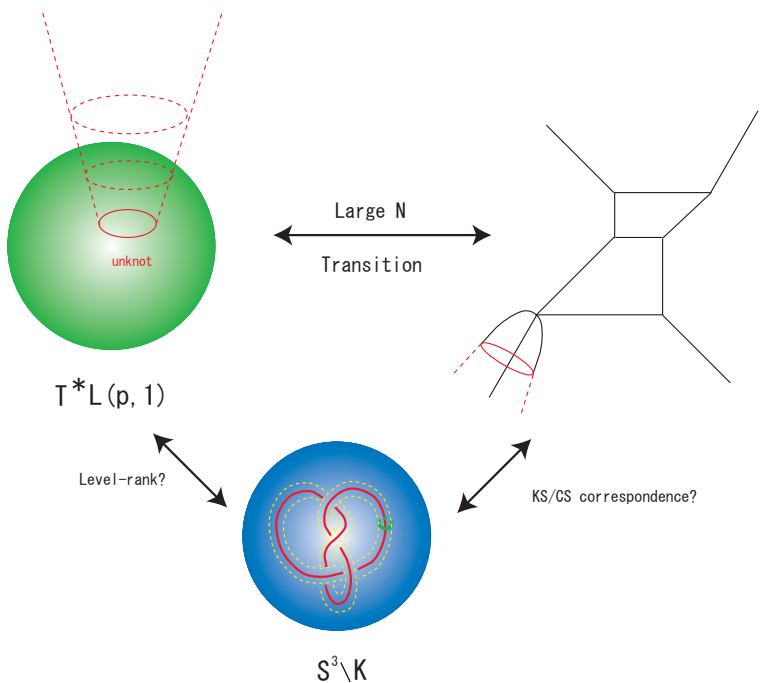}
\caption{Large $k$ duality via level-rank duality.}
\end{figure}

It will also be important to study such dualities by embedding the
hyperbolic three-manifold directly in the non-compact Calabi-Yau
threefold as the cotangent bundle.  The $U(2)$ Chern-Simons gauge
theory on the hyperbolic three-manifold $M$ is realized as the
topological A-model on $T^*M$ which the two compact D-branes
wrap around $M$.  However, in the study on the cotangent bundle
\cite{Feix}, it is shown that there are no complete Ricci-flat
K\"ahler metrics on $T^*M$.  We do not know how this theorem
invalidates the stringy realization of the hyperbolic geometry. The
subtlety of this point should be considered in more detail.

Arithmetically, the {\it Ruelle $L$-function} is defined 
for the hyperbolic three-manifold.
\begin{eqnarray}
R_{M}(z,\rho):=\prod_{\gamma}\det [1-\rho(\gamma)e^{-zl(\gamma)}],
\end{eqnarray}
where $l(\gamma)$ is the length of $\gamma$ and 
$\gamma$ runs through the primitive closed geodesics.
The leading term in the expansion of $z$ around the origin 
coincides with the volume of the hyperbolic three-manifold $M$ 
\cite{Mathai,Lott}.
The subleading term of the Ruelle $L$-function
is \cite{Park,Sugiyama}
\begin{eqnarray}
R_M(0,\rho)=\tau(M,\rho)^2,
\end{eqnarray}
where $\tau(M,\rho)$ is the Franz-Reidemeister torsion. In particular,
for the fibered knots, the subleading term
coincides with the twisted Alexander polynomial \cite{Kirk,Kitano,Wada}.
For the fibered knot complement, the twisted Alexander polynomial 
is computed explicity in terms of the variable of the character variety 
\cite{Dubois2}. 
The result coincides with the torsion $T(M;\rho)$ for not the knot 
complement but its Dehn-surgered manifold.\footnote{
In the notation of \cite{Porti}, the twisted Alexander polynomial
coincides with not $TOR(C_0)$ but $TOR(C_1)$.}
Although there exist such subtle points, we expect some equivalence 
between the Ruelle $L$-function and the topological open string partition 
function, and such a relation may clarify the arithmetic aspects of 
our proposal.

Recently, the asymptotic analysis of the colored Alexander invariants
\cite{ADO} has been carried out \cite{col_alex}. The colored Alexander
invariant is closely related with the logarithmic conformal field theory
and is useful for studying the link invariants. In \cite{col_alex}, the
A-polynomial for the link is also obtained by computing an analogue of
the Neumann-Zagier function \cite{Murakami-Yokota} from the
asymptotics of the colored Alexander polynomial.  It is interesting to
consider the interpretation of the new volume conjecture in terms of the
topological string theory.

In \cite{DVKS}, the lift of the Kodaira-Spencer theory on the Riemann
surface to the three-dimensional Chern-Simons gauge theory is proposed. In
our setup, we considered the Kodaira-Spencer theory on the character
variety $X_K$, and the corresponding Chern-Simons gauge theory should
be defined on $X_K\times\mathbb{R}$.  At present, we cannot find
the relation between $X_K\times\mathbb{R}$ and the knot complement
${\bf S}^3\backslash K$, but the Chern-Simons gauge
theories on both of these three-manifolds may be related via some
surgeries or analytic continuation of the flat connections.

\vspace{1cm}
\noindent{\bf Acknowledgements:}

We would like to thank T. Eguchi, B. Eynard, D. Jadnanansing,
R. Kashaev, A. Kashani-Poor and  R. van der Veen for fruitful
discussions and useful comments.
One of the authors (H.F.) thanks to H. Awata, A. Brini, K. Hikami,
K. Hosomichi, K. Ito, M. Jinzenji, H. Kanno, A. Kato, M. Kurachi,
M. Manabe, T. Masuda, S. Mizoguchi, H. Murakami, J. Murakami,
T. Nakatsu, Y. Sasai, N. Sasakura, M. Shigemori, H. Suzuki, 
S. Terashima, T. Tokunaga, K. Tsuda, J. Walcher, M. Yamazaki, and N. Yotsutani 
for discussions and encouragements. 
H.F. is also grateful to Institute for Theoretical Physics, 
University of Amsterdam
and SISSA/ISAS for warm hospitality.  
The work of H. F. 
is supported by the Grant-in-Aid for Nagoya
University Global COE Program, ``Quest for Fundamental Principles in the
Universe: from Particles to the Solar System and the Cosmos'', 
from the Ministry of Education, Culture, Sports, Science and Technology of Japan,
and 
the research of R.D. is supported by a NWO Spinoza grant and the FOM
program {\it String Theory and Quantum Gravity}.


\appendix

\section*{Appendix~~ Derivation of annulus formula}
\setcounter{equation}{0} \def\theequation{A.\arabic{equation}} 

In this appendix, we summarize the derivation of the annulus formula for the matrix
model \cite{FM}.  The partition function of the Chern-Simons matrix
model on the Lens space is given by \cite{Marino,AKMV}
\begin{eqnarray}
Z(L(p,1))=\int\prod_{i=1}^{N}du_i \prod_{j<k}2\sinh^2\frac{u_j-u_k}{2}
\exp\left[-\frac{p}{g_s}\sum_{i=1}^N \left(\frac{u_i^2}{2}
-\frac{2\pi i}{p}n_i u_i \right)\right],
\label{CS matrix}
\end{eqnarray}
where $n_i=I$ for $N_{I-1}\le i<N_I$ and $\sum_{I=1}^p N_I=N$.
The resolvent operator for this matrix model is
\begin{eqnarray}
&&\omega(z):=\sum_{I=1}^p\omega_I\left(z-\frac{2\pi iI}{p}\right), \quad
\omega_I(z):=g_s\sum_{i=1}^{N_I}\coth\frac{z-u_i}{2}.
\end{eqnarray}
In the large $N$ limit, this resolvent operator satisfies the loop equation
\label{HOY}
\begin{eqnarray}
&&(e^{v}-1)(e^{pz}e^v-1)+e^S-1+e^v\sum_{n=1}^{p-1}d_ne^{nz}=0, \\
&& v:=\frac{S}{2}-\frac{\omega}{2}, \quad S:=g_s N.
\end{eqnarray}

The vacuum expectation value $W_R$ of the Wilson loop operator 
for the $U(N)$ Chern-Simons gauge theory along an unknot in $L(p,1)$ 
with the representation $R$ is also given in terms of the Chern-Simons 
matrix model \cite{remodel}
\begin{eqnarray}
&&W_R=\frac{1}{Z(L(p,1))}\int\prod_{i=1}^Ndu_i \prod_{j<k}2\sinh^2\frac{u_j-u_k}{2}
\exp\left[-\frac{p}g_s\sum_{i=1}^{N}\left(\frac{u_i^2}{2}-\frac{2\pi i}{p}n_iu_i\right)
\right]{\rm Tr}_R U,
\nonumber \\
&&U:={\rm diag} (e^{-u_1},e^{-u_2},\cdots,e^{-u_N}).
\end{eqnarray}
The D-brane partition function on the non-compact Calabi-Yau threefold,
which is a mirror to
\begin{eqnarray}
X^{\vee}:=\Bigl\{(x,y,z,w)\in (\mathbb{C}^*)^2\times \mathbb{C}^2
\Bigl|
zw=(y-1)(x^p y-1)+e^S-1+y\sum_{n=1}^{p-1}d_nx^n\Bigr\},
\end{eqnarray}
is given by $W_R$. 
\begin{eqnarray}
&&Z_D(a)=\sum_R{W_R}{\rm Tr}_R e^{-a}
\nonumber \\
&=&\frac{1}{Z(L(p,1))}\int \prod_{i=1}^N du_i
\prod_{j<k}2\sinh^2\frac{u_j-u_k}{2}
\exp\left[
-\frac{p}{g_s}\sum_{i=1}^N\left(\frac{u_i^2}{2}
-\frac{2\pi i}{p}n_iu_i\right)
\right]
\nonumber \\
&&\quad\quad\quad\quad\quad\quad \times
\sum_{R}s_R(e^{-u_1},\cdots,e^{-u_N})s_R(e^{-a})
\nonumber \\
&=& \langle \det (\mathbb{I}_N-e^{-a}U)^{-1}\rangle_{N}
\nonumber \\
&=&
\int\prod_{i=1}^{N}du_i\prod_{j<k}2\sinh^2\frac{u_j-u_k}{2}
\exp\left[-\frac{p}{g_s}\sum_{i=1}^N\left(
\frac{u_i^2}{2}-\frac{2\pi i}{p}n_iu_i+\frac{g_s}{p} 
\log(1-e^{-a}e^{-u_i})
\right)\right],
\nonumber \\
&&
\label{CS matrix2}
\end{eqnarray}
where $s_R(x)$ denotes a Schur function \cite{Mac}.  By comparing
(\ref{CS matrix}) and (\ref{CS matrix2}), we can introduce the open
string expansion parameter $\epsilon$, which counts the number of holes
as follows:
\begin{eqnarray}
Z_D(a;g_s,\epsilon)&:=&\frac{1}{Z(L(p,1))}\int\prod_{i=1}^N du_i
\prod_{j<k}2\sinh^2\frac{u_j-u_k}{2}
\nonumber\\
&&\times
\exp\Bigl[
-\frac{p}{g_s}\sum_{i=1}^N\Bigl(\frac{u_i^2}{2}
-\frac{2\pi i}{p}n_iu_i+\frac{\epsilon}{p} 
\log(1-e^{-a}e^{-u_i})
\Bigr)
\Bigr].
\end{eqnarray}

In the Chern-Simons matrix model (\ref{CS matrix}), there can exist
$p$ cuts around the critical points $u=\frac{2\pi i}{p}I$ 
($I=1,\cdots, p$).  The saddle point equation around the $I$-th cut $C_I$
is
\begin{eqnarray}
p\left[u_i-\frac{2\pi i}{p}I\right]=g_s\sum_{j\ne
 i}\coth\frac{u_i-u_j}{2}, \quad u_i\in C_I.
\label{saddle1}
\end{eqnarray}
When the D-brane is introduced, we solve the saddle point
equation for $Z_D(a;g_s,\epsilon)$ as
\begin{eqnarray}
p\left[u_i-\frac{2\pi i}{p}I+\frac{\epsilon}{p}
\frac{e^{-a}e^{-u_i}}{1-e^{-a}e^{-u_i}}\right]
=g_s\sum_{j\ne i}\coth\frac{u_i-u_j}{2}, \quad u_i\in C_I.
\label{saddle2}
\end{eqnarray}
In the continuum limit $N\to \infty$, we obtain
\begin{eqnarray}
-\frac{p}{S}\left[
u(s_0)-\frac{2\pi i}{p}\sum_{I=1}^p\theta(s_0-\frac{N_I}{N})+\frac{\epsilon}{p}\frac{e^{u(s_0)}}{e^{u(s_0)}-e^{-a}}
\right]=-\hspace{-0.45cm}\int_0^1 ds\coth\frac{u(s)-u(s_0)}{2},
\label{saddle3}
\end{eqnarray}
where $u_i$ is replaced by a continuous function $u(s)$ on $0\le
s\le1$.  Here, we define the eigenvalue density $\rho({\cal U})$
\begin{eqnarray}
ds=\frac{d{\cal U}}{{\cal U}}\rho({\cal U}),\quad {\cal U}:=e^{u(s)},
 \quad {\cal U}_0:=e^{u(s_0)},\quad
\int_{C_I}\rho({\cal U})\frac{d{\cal U}}{{\cal U}}=\frac{N_I}{N}.
\end{eqnarray}
Using this eigenvalue density, we can rewrite the saddle point equation
(\ref{saddle3}) as
\begin{eqnarray}
-\frac{p}{2S}\left[
\log ({\cal U}_0e^{-S/p})-\frac{2\pi i}{p}\sum_{I=1}^p
\theta(s_0-\frac{N_I}{N})+\frac{\epsilon}{p}
\frac{{\cal U}_0}{{\cal U}_0-{\cal A}}\right]
=-\hspace{-0.45cm}\int_C
d{\cal U}\frac{\rho({\cal U})}{{\cal U}-{\cal U}_0},
\end{eqnarray}
where ${\cal A}:=e^{-a}$. The eigenvalue density $\rho({\cal U})$ and
the resolvent $\omega(u;\epsilon)$ are related on each cut $C_I$.
\begin{eqnarray}
&& -2\pi i\rho({\cal U})=\omega(u+i0;\epsilon)-\omega(u-i0;\epsilon),
\nonumber \\
&&
\frac{1}{2}[\omega(u+i0;\epsilon)+\omega(u-i0;\epsilon)]
=pu+\epsilon\frac{{\cal U}}{{\cal U}-{\cal A}}-2\pi iI.
\end{eqnarray}
These relations imply 
\begin{eqnarray}
\omega(z\pm i0;\epsilon)=-\frac{p}{2S}\left[
\log({\cal Z}e^{-S/p})-\frac{2\pi i}{p}I+\frac{\epsilon}{p}\frac{{\cal
Z}}{{\cal Z}-{\cal A}}
\right]\mp \pi i\rho({\cal Z}), \quad z\in C_I,
\end{eqnarray}
where ${\cal Z}:=e^{z}$.
Therefore, the resolvent $\omega(z)$ satisfies the following conditions:
\begin{description}
\item{{\it (i)}} Integrating around each cut $C_I$, we obtain the
	   following filling fraction:
\[
\frac{1}{2\pi i}\oint_{C_I}dz\;\omega(z)=\frac{N_I}{N}.
\]
\item{{\it (ii)}} $\omega(z)$ decays at infinity as 
$\omega(z)dz\sim \frac{d{\cal Z}}{\cal Z}+{\cal O}(\epsilon)$.
\item{{\it (iii)}} $\omega(z)dz$ does not have any simple poles except
	   for the infinity on the first sheet.
\item{{\it (iv)}} $\omega(z)dz$ has  simple poles at ${\cal Z}={\cal A}$
	   and $\infty$ on the second sheet with the residues
	   $\epsilon\frac{1}{S}$ and $-\epsilon\frac{1}{S}$,
	   respectively.
\end{description}

From these conditions, the resolvent is determined uniquely up to ${\cal
O}(\epsilon)$.
\begin{eqnarray}
&& \omega({\cal Z};\epsilon)
=\omega^{(0)}({\cal Z})+\epsilon\omega^{(1)}({\cal
 Z})+{\cal O}(\epsilon^2).\\
&& \omega^{(0)}({\cal Z})=\frac{1}{S}\log {\cal Y}({\cal Z}), 
\label{omega0}\\
&& \omega^{(1)}({\cal Z})d{\cal Z}=\frac{1}{S}\omega_{\tilde{\cal
 A}-\tilde{\infty}}.
\label{omega1}
\end{eqnarray} 
Here, ${\cal Y}({\cal Z})$ is a solution of the loop equation which 
converges as ${\cal Z}\to\infty$.
\begin{eqnarray}
{\cal Z}^p{\cal Y}^2+(-{\cal Z}^p+\sum_{n=1}^{p-1}d_n{\cal Z}^n-1){\cal
 Y}+e^S=0. \label{spec3}
\end{eqnarray}
$\tilde{A}$ and $\tilde{\infty}$ are points at ${\cal Z}={\cal A}$ and
$\infty$ on the second sheet in the covering of the spectral curve
(\ref{spec3}).  $\omega_{\tilde{\cal Z}_1-\tilde{\cal Z}_2}$ is an
Abelian differential of the third kind with simple poles at
$\tilde{\cal Z}_1$ and $\tilde{\cal Z}_2$ with residues $+1$ and $-1$,
respectively, and zero A-periods.

Using this large $N$ solution, we can compute the planar free energy.
The free energy is expanded in terms of $g_s$ and $\epsilon$ as
\begin{eqnarray}
&& Z_D(a;g_s,\epsilon)
=\exp[\sum_{g=0}^{\infty}g_s^{-2}F(a;g_s,\epsilon)], 
\\
&&
F(a;g_s,\epsilon)=\sum_{g=0}^{\infty}g_s^{2g}F_g(a;\epsilon)
=\sum_{g=0}^{\infty}\sum_{h=1}^{\infty}g_s^{2g}\epsilon^h 
F_{g,h}(a).
\end{eqnarray}
In particular, for the planar topologies $g=0$, 
the free energy  yields to
\begin{eqnarray}
F_{0,h}&=&\frac{1}{h!}\frac{\partial^h F_0(\epsilon)}{\partial
 \epsilon^h}\Bigg|_{\epsilon=0}
=\frac{1}{h!}\lim_{g_s\to 0}\frac{\partial^h F(a;g_s,\epsilon)}{\partial
 \epsilon^h}\Bigg|_{\epsilon=0}
\nonumber \\
&=& \frac{1}{h!}\lim_{g_s\to 0}\frac{\partial^{h-1}}{\partial
 \epsilon^{h-1}}\left(-g_s
\biggl\langle
{\rm Tr}_N\log(\mathbb{I}_N-e^{-a}U)
\biggr\rangle
\right),
\end{eqnarray}
where the $g_s\to 0$ limit implies the  
 't Hooft limit, which fixes the
value $S=g_sN$.  In this limit, the one-point function $\biggl\langle
{\rm Tr}_N\log(\mathbb{I}_N-e^{-a}U)\biggr\rangle$ can be computed by
utilizing the eigenvalue density $\rho({\cal U})$.
\begin{eqnarray}
&& \lim_{g_s\to 0}\frac{1}{N}\langle
{\rm Tr}_N
\log(\mathbb{I}_N-e^{-a}U)
\rangle_{g_s,\epsilon}
=\int_{-\infty}^{\infty}d{\cal U}\rho({\cal U})
\log(1-{\cal A}{\cal U}^{-1})
\nonumber \\
&&=\int_{-\infty}^{\infty}d{\cal U}\rho({\cal U})
\log({\cal U}-{\cal A})
+\int_{-\infty}^{\infty}d{\cal U}\rho({\cal U})\log {\cal U},
\nonumber \\
&&=-\frac{1}{2\pi i}\sum_{I=1}^p\oint_{C_I}d{\cal Z}\;
\omega({\cal Z};\epsilon)\log({\cal Z}-{\cal A})+c_0.
\label{free}
\end{eqnarray}
where $c_0:=\int_{-\infty}^{\infty}d{\cal U}\rho({\cal U})\log {\cal U}$.
In the following discussion, we neglect this constant factor.

By applying the large $N$ solutions (\ref{omega0}) and (\ref{omega1}) to
(\ref{free}), we can compute the disk and the annulus free energies.
\begin{eqnarray}
&& F_d(a)=-\frac{1}{2\pi i}\sum_{I}\oint_{C_I}d{\cal Z}\;\log{\cal Y}({\cal
 Z})\log({\cal Z}-{\cal A})=\int_{\cal A}^{\infty}d{\cal Z}\;\log {\cal
 Y}({\cal Z}), \\
&& F_a(a)=-\frac{1}{4\pi i}\sum_{I}\oint_{C_I}\omega_{\tilde{\cal A}-\tilde{\infty}}\log({\cal Z}-{\cal A})=\frac{1}{2}\int_{\cal A}^{\infty}\omega_{{\tilde{A}-\tilde{\infty}}}.
\end{eqnarray}
The free energy $F_d(a)$ for the disk topology
is consistent with the result of the
B-model computation (\ref{rhCS}).  In contrast, the free
energy $F_a(a)$ for the annulus topology can be rewritten further as
\begin{eqnarray}
F_a(a)=\frac{1}{2}\log
\frac{E({\cal A},\tilde{\infty})E(\infty,\tilde{\cal A})}
{E({\cal A},\tilde{\cal A})E(\infty,\tilde{\infty})},
\end{eqnarray}
where $E({\cal Z},{\cal W})\sqrt{d{\cal Z}}\sqrt{d{\cal W}}$ is the prime
form on the spectral curve (\ref{spec3}).  A similar result is
obtained in topological string computations \cite{remodel}.

\end{document}